\documentclass[prx,aps,showpacs,twocolumn,preprintnumbers,amsmath,amssymb,superscriptaddress,longbibliography]{revtex4-2}
\usepackage{color,graphicx,pstricks,float}
\usepackage{natbib}
\usepackage{fancybox}
\usepackage{physics}
\usepackage{epsfig}
\usepackage{graphicx}
\usepackage{tabularx}
\usepackage{inputenc}
\usepackage{amsmath,braket,amsfonts,amsmath,mathtools,mathrsfs}
\usepackage{amssymb}
\usepackage{gensymb}
\usepackage{grffile}
\usepackage{relsize}
\usepackage{array}
\usepackage{comment}
\usepackage{amssymb}
\usepackage{hyperref}
\usepackage[table,dvipsnames]{xcolor}
\usepackage{gensymb}
\usepackage{grffile}
\usepackage{float}
\usepackage{bibunits}
\usepackage{orcidlink}
\hypersetup{
colorlinks=true,
linkcolor = blue,
filecolor = blue,
urlcolor = Blue,
citecolor = blue,
pdftitle = {}
}

\newcommand{\mk}{\mathbf{k}}

\newcommand{\q}[1]{``#1''}
\begin{document}

\title{Quantum geometric anomalous Hall response in orbitally nonunitary superconductors}
\author{Viktor Frilén,\,\orcidlink{0009-0004-0251-5953}}
\affiliation{Department of Physics and Astronomy, Uppsala University, Box 524, SE 751 20 Uppsala, Sweden}
\affiliation{Department of Physics, Gothenburg University, 41296 Gothenburg, Sweden}
\author{Annica M. Black-Schaffer\,\orcidlink{0000-0002-4726-5247}}
\author{Ankita Bhattacharya,\,\orcidlink{0000-0002-7432-9533}}
\email{ankita.bhattacharya@physics.uu.se}
\affiliation{Department of Physics and Astronomy, Uppsala University, Box 524, SE 751 20 Uppsala, Sweden}
%\affiliation{Department of Physics, Gothenburg University, 41296 Gothenburg, Sweden}

\begin{abstract}
We investigate the anomalous Hall response (AHR) in a multiband superconductor at optical frequencies, a phenomenon intimately related to the polar Kerr effect, a key probe of time-reversal symmetry breaking in superconductors. In translationally invariant multiband systems with purely intraband pairing, Galilean invariance decouples center-of-mass and relative motion of Cooper pairs, leading to the widespread expectation that a finite AHR requires either disorder or finite interband pairing amplitudes. However, this restriction can be lifted by the quantum geometric effects inherent to multiband Bloch states. Using a honeycomb lattice tight-binding model with Kane–Mele spin–orbit coupling, we analyze the AHR for the time-reversal symmetry broken chiral $d$-wave spin-singlet and chiral $p$-wave equal-spin-triplet pairing states with \emph{intraband} pairing only. We demonstrate, through both analytical and numerical calculations that the spin-singlet state yields a vanishing AHR, even with its broken time-reversal symmetry, whereas the equal-spin triplet state exhibits a finite AHR, even when it is spin-unitary. We attribute the latter to orbital nonunitarity, which, in the presence of spin–orbit coupling, generates the spin-polarized Bogoliubov quasiparticle states required for a finite AHR. The response is mediated by interband velocity matrix elements governed by the quantum geometry. This finding establishes that spin-unitary, but orbitally nonunitary pairing, can generate a finite AHR even without interband pairing and thereby revises the criteria for Kerr signals in superconductors.
\end{abstract}

\date{\today}
\maketitle
\textcolor{blue}{\textit{Introduction.}} The anomalous Hall response (AHR) in superconductors at optical frequencies is generating the polar Kerr effect, in which reflected light exhibits a finite rotation of its plane of polarization.  
The experimentally observed finite Kerr angle in superconductors, such as  $\mathrm{Sr_2RuO_4}$~\cite{Kapitunik06}, $\mathrm{UTe}_2$~\cite{Hayes21} and $\mathrm{UPt_3}$~\cite{Schemm14}, has in turn been providing a key signature for time-reversal symmetry (TRS) breaking in the superconducting state \cite{Kapitulnik_2009}.

While TRS breaking is a necessary condition for finite AHR in both normal metals and superconductors~\cite{Nagaosa10,Xiao10}, the realization of AHR in superconductors is far more constrained. Unlike normal metals, where transport is carried by electrons in Bloch bands, the low-energy response of a superconductor is governed by a phase-stiff condensate and Bogoliubov quasiparticles, leading to a qualitatively different electromagnetic response under an external electric field. As a consequence, TRS breaking is not sufficient to generate a finite AHR in superconductors, for example, in a clean single-band superconductor with TRS breaking pairing~\cite{Roy08,Lutchyn08} due to Galilean invariance~\cite{Read2000}. The TRS breaking resides in the internal (relative coordinate) structure of Cooper pairs, whereas the external electric field couples only to their center-of-mass (CM) motion, thereby not generating a finite AHR. 

Several proposals have been put forward to couple the relative and CM motion of Cooper pairs to generate a finite AHR for TRS breaking pairing, either \emph{extrinsically} through disorder-induced breaking of translational invariance~\cite{Goryo08,Lutchyn09,Levchenko17,Li20,Liu23} or \emph{intrinsically} by interband pairing in clean multiband superconductors~\cite{kallin12,Mineev12,Gyroffy12,Kallin13}. More recent theoretical works have demonstrated that an intrinsic AHR can also emerge in multiband superconductors even in the absence of interband pairing ~\cite{Brydon21,Zhang24}. 
Here the relative and CM motion are instead coupled through the \emph{interband} velocity, the off-diagonal element of the velocity matrix in the band basis~\cite{Zhang24,Hu25,Bhattacharya26}. Notably, the interband velocity is closely connected to the quantum geometric properties of the Bloch bands~\cite{Torma23}, thereby providing an intrinsic geometric origin for the AHR. 
Existing realizations of the AHR arising from intraband pairing have required not only a finite spin-orbit coupling (SOC) in the normal state but also nonunitary spin-triplet pairing, which induce the spin imbalance~\cite{Brydon21,Zhang24} required for a finite AHR.

However, the notion of nonunitarity of the pairing in multiband systems is intricate, as it can originate from both spin or orbital (sublattice) degrees of freedom, or sectors. Consequently, a spin-singlet state that is unitary in the spin sector can nevertheless exhibit nonunitarity due to contributions from the orbital sector~\cite{Zeng23}. In fact, it has been shown that, in the presence of SOC in the normal state, even unitary spin pairing can exhibit a spin-imbalance when nonunitarity arises from another pairing sector~\cite{Zeng23}.  An intriguing question thus arises whether a spin-unitary, yet globally nonunitary, superconducting pairing state may generate finite AHR, and consequently a finite Kerr signal, in the presence of intraband pairing only? This is the central question we address in this work.

In this Letter, we use a tight-binding honeycomb model, with symmetry-allowed Kane-Mele SOC~\cite{KM05,KM052nd}, to investigate the AHR of two prototype TRS breaking superconducting states, which also host orbital nonunitarity ~\cite{Brydon19,Faye15}: the chiral $d$-wave spin-singlet and the chiral $p$-wave equal-spin-triplet state. While the spin-singlet state is always unitary in the spin sector, the equal-spin-triplet state can be either unitary or nonunitary~\cite{Ramires_2022}, however, in both cases, the orbital sector is generally nonunitary. Through both analytical derivations and numerical simulations, we show that the spin-singlet state cannot support any AHR with only intraband pairing, while the equal-spin-triplet state can exhibit a finite AHR, notably even for only intraband pairing and when it remains unitary in the spin sector. The latter is due to the combination of spin-orbit coupling and orbital nonunitarity producing a finite spin-imbalance, and mediated by quantum geometry.
These results explicitly illustrate how the interplay between quantum geometry, and spin and orbital degrees of freedom, specifically, nonunitary components of the orbital sector, can give rise to a finite AHR, even in the absence of interband pairing.
Our results open up exciting new possibilities for identifying and understanding TRS-breaking phenomena in superconducting systems, even in the absence of interband pairing. Moreover, our results offer a compelling new direction for future experimental investigations, where the role of quantum geometry and the subtle interplay between spin, orbital, and pairing symmetries can be probed in a more systematic way. 
%%%%%%%%%%%%%%%%%%%%%%%%%%%%%%%%%%%%%%%%%%%%%%%%%%%%%%%%%%%%%%%%%%%%%%%%%%%%%%%%%%%%%%%%%%%%%%%%%%%%%%%%%%%

\noindent
\textcolor{blue}{\textit{Model}.} To investigate the intrinsic anomalous Hall conductivity in a multiband superconductor, we use a tight-binding model on the honeycomb lattice for the normal-state electronic structure~\cite{KM05,KM052nd}
\begin{equation}
\resizebox{\columnwidth}{!}{$
H_N =  t \sum_{\!\langle i,j\rangle, \alpha} c_{i,\alpha}^\dagger c_{i,\alpha}
    - \mu \sum_{\!i,\alpha\!} c_{i,\alpha}^\dagger c_{i,\alpha} 
  + i \lambda_{\text{SOC}}\!\!\sum_{\langle\!\langle i,j\rangle\!\rangle, \alpha \beta\!}
    \nu_{ij}\, \sigma_z^{\alpha\beta} c_{i,\alpha}^\dagger c_{i,\beta},
$}
\label{eq:normal state Hamiltonian}
\end{equation}
where $ c_{i,\alpha}^\dagger,  c_{i,\alpha}$ are the creation and annihilation operators of electrons at the sublattice site $i$ with spin $\alpha$. The symbols $\langle \cdot\cdot\cdot\rangle$, $\langle\langle \cdot\cdot\cdot\rangle\rangle$ denote summations over the nearest and next-nearest neighbors, respectively, and $\sigma_i$ is a Pauli matrix in spin space. Here, $\mu$ is the chemical potential, $\lambda_{\text{SOC}}$ denotes the symmetry-allowed Kane-Mele type SOC term, and $\nu_{ij}=-\nu_{ji}=\pm 1$ specifies clockwise and anticlockwise hopping within each hexagonal plaquette. $H_N$ preserves both inversion symmetry and TRS, and hosts additional degrees of freedom due to the two sublattices, hereafter referred to as the orbital degree of freedom. 

To study superconductivity, we consider the Bogoliubov-de Gennes (BdG) Hamiltonian 
\begin{align}
\mathcal{H}_{\text{BdG}} (\mathbf{k}) = \sum_{\mathbf{k}} 
\Psi_{\mathbf{k}}^\dagger
\begin{pmatrix}
H_N(\mathbf{k}) & \Delta(\mathbf{k}) \\
\Delta^\dagger(\mathbf{k}) & -H_N^T(\bar{\mathbf{k}})
\end{pmatrix}
\Psi_{\mathbf{k}},
\label{eq:BdG Hamiltonian}
\end{align}
where $\Psi^\dagger_\mk = (\psi^\dagger_\mk, \psi^T_{-\mk})$ is the Nambu spinor and $\psi_\mathbf{k}=(c^\dagger_{\mk A \uparrow}, c^\dagger_{\mk B \uparrow}, c^\dagger_{\mk A \downarrow}, c^\dagger_{\mk B \downarrow})$ with $A, B$ being sublattice indices. Here $H_N(\mathbf{k})$ is obtained by Fourier transformation of the Hamiltonian in Eq.~(\ref{eq:normal state Hamiltonian}), see Supplemental Material (SM)~\cite{SM} and we use $\bar{\mathbf{k}}=-\mathbf{k}$. 
The $4\times 4$ pairing matrix $\Delta(\mk)$ defines the superconducting state. Electron-electron interactions in the honeycomb lattice can favor Cooper pairing with nonzero angular momentum. In particular, both the TRS breaking spin-singlet chiral $d$-wave and spin-triplet chiral $p$-wave states have been proposed~\cite{BS07, Bhaskaran10,BS2014,Meng16,Wang24}. Motivated by these results, we here consider both of these pairing states.

The simplest chiral states in sublattice space takes the form~\cite{Brydon19,Faye15}
\begin{align}
    \Delta^\eta_\pm(\mk) = \Delta_0 \sum_{j=1}^3 e^{\mp \eta i\phi_j} 
    \begin{pmatrix}
        0 & e^{i\mk \cdot \mathbf{R}_j} \\
        \eta e^{-i\mk \cdot \mathbf{R}_j} & 0
    \end{pmatrix}, \label{eq:orbital_pairing_matrix}
\end{align}
where $\eta = + 1$ $(-1)$ for $d$\,($p$)-wave order parameters and $\pm$ denotes the two degenerate chiral states, related by TRS. This is nearest-neighbor pairing with uniform amplitude $\Delta_0$ and a bond-dependent phase difference $\phi_j = (j-1)2\pi/3$, with $j=1,2,3$, and $\mathbf{R}_j$ nearest-neighbor bond vectors. 

The $d$-wave order parameter can be decomposed in terms of the basis functions of the two-dimensional irreducible representation (irrep) $E_{2g}$ of the lattice point group $D_{6h}$, such that $\Delta^d_\pm(\mk) = \Delta^d_{x^2-y^2}(\mathbf{k}) \pm i \Delta^d_{xy}(\mk)$. Similarly, the $p$-wave order parameter is decomposed in terms of the irrps of the two-dimensional $E_{1u}$ irrep: $\Delta^p_\pm (\mathbf{k}) = \Delta_x^p(\mathbf{k}) \pm i\Delta^p_y(\mathbf{k})$. These complex combinations reflect the inherent chiral nature of superconductivity that breaks TRS.  By including the spin degrees of freedom, the full pairing matrix can be written as
\begin{align}
    \Delta(\mk) = \Delta_{\text{spin}}\otimes \Delta_{\text{orb}} = \Delta_{\text{spin}}\otimes \Delta_\pm ^\eta(\mk),\label{eq:pairing_def}
\end{align}
where $ \Delta_{\text{spin}}$ corresponds to a spin-singlet pairing state for chiral $d$-wave ($\eta=1$)  and a triplet pairing state for chiral $p$-wave  ($\eta=-1$). This decomposition into orbital and spin sectors remain valid even in the presence of the Kane-Mele SOC, since this SOC preserves $S_z$ conservation.

For the chiral $d$-wave spin-singlet pairing, $\Delta_\text{spin} = i \sigma_y$, whereas for the chiral $p$-wave pairing, we choose to consider an equal-spin-triplet pairing, for which the spin-sector pairing matrix takes the form
\begin{align}
  \Delta_\text{spin} & =  \begin{pmatrix}
        \Delta^{\uparrow\uparrow} & 0 \\
        0 & \Delta^{\downarrow\downarrow}
    \end{pmatrix}= \begin{pmatrix}
      1 & 0 \\
        0 & \kappa
    \end{pmatrix}   \Delta^{\uparrow\uparrow}, \label{eq:delta_spin}
\end{align}
where $\kappa =  \Delta^{\downarrow\downarrow}/\Delta^{\uparrow\uparrow}$ controls the relative pairing amplitude, and the remaining amplitude $\Delta^{\uparrow\uparrow}$ is absorbed in the overall amplitude $\Delta_0$ in Eq.~(\ref{eq:orbital_pairing_matrix}). 
\begin{figure}[t!]
    \centering
    \includegraphics[width=0.8\columnwidth]{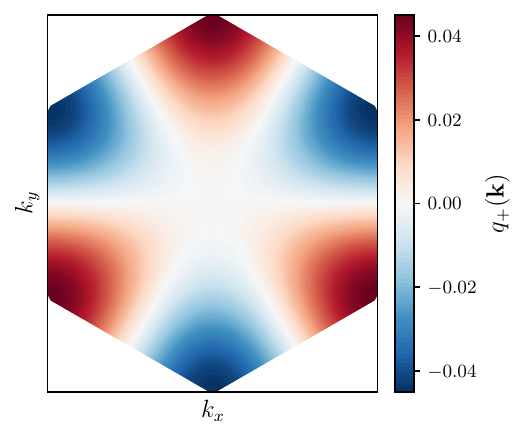}
    \caption{Momentum-space distribution of $q_\pm$ in Eq.(\ref{eq:orbital_nonunitary_def}) for the chiral d-wave ($\eta=1$) spin-singlet pairing state for $\Delta_0=0.1 t$. }
    \label{fig:Fig3}
\end{figure}

\noindent
\textcolor{blue}{\textit{Nonunitarity pairing}.}
A pairing state is termed nonunitary, when the pairing matrix $\Delta(\mathbf{k})$ in Eq.~(\ref{eq:BdG Hamiltonian}) satisfies $\Delta(\mathbf{k}) \Delta(\mathbf{k})^\dagger \not \propto I$~\cite{Ramires_2022}.
In a simple single-band system, nonunitarity can only arise  in for spin-triplet pairing. In contrast, in a multiband system, such as Eq.~(\ref{eq:pairing_def}), the pairing matrix can be nonunitary if either the orbital part $\Delta_\pm ^\eta(\mk)$ or the spin part $\Delta_{\text{spin}}$ or both is nonunitary. Consequently, even a spin-singlet state may be nonunitary, provided that the orbital (or sublattice) pairing sector is nonunitary~\cite{Zeng23}.

For spin-singlet pairing, the spin pairing matrix is always unitary, as $\Delta_\text{spin} \Delta^\dagger_\text{spin} \propto \sigma_0$. For the spin-triplet pairing states in Eq.~(\ref{eq:delta_spin}) we get 
\begin{align}
    \Delta_\text{spin} \Delta^\dagger_\text{spin} = 2 (1 + \kappa^2) \sigma_0 + 2 (1 - \kappa^2) \sigma_3, 
\end{align}
which indicates that the spin pairing matrix is unitary if $\kappa=\pm 1$, but nonunitary otherwise. Further, any deviation of $\kappa$ from unity means unequal pairing between the two spin channels. This leads to a finite net spin polarization (spin imbalance) of the Cooper pair, which can thus be viewed as a hallmark of nonunitarity. 

Similarly, to check the unitarity condition for the pairing matrix in the orbital sector, we examine the matrix product $\Delta^\eta_{\pm} \Delta^{\eta\dagger}_{\pm}$, which takes the form
\begin{align}
   \Delta^\eta_{\pm} \Delta^{\eta\dagger}_{\pm} = \Delta_{U} \,\tau_0 + q_{\pm} \tau_3, \label{eq: Deltao_nonunitary}
\end{align}
with
\begin{align}
    \Delta_U &= |\Delta_0|^2 \Big[1 + 2 \sum_{j>l} \cos(\phi_j-\phi_l)\cos(\mk\cdot[\mathbf{R}_j-\mathbf{R}_l]) \Big], \\
    q_{\pm} &= (\pm \eta)\, 2|\Delta_0|^2 \sum_{j>l} \sin(\phi_j-\phi_l)\sin(\mk\cdot[\mathbf{R}_j-\mathbf{R}_l]),
    \label{eq:orbital_nonunitary_def}
\end{align}
where $\tau$ acts in the sublattice (orbital) space. Analogous to the condition for nonunitarity in spin space, the pairing matrix in orbital space becomes nonunitary only when $q_{\pm}\neq 0$. 
We find a finite $q_\pm$ in most of the Brillouin zone, see Fig.\ref{fig:Fig3}. 
While nonunitarity in the orbital sector by itself gives no spin imbalance, adding SOC has been shown to give a finite spin-polarization, also for spin-singlet pairing~\cite{Zeng23}.
\begin{figure}
    \centering
    \includegraphics[width=\columnwidth]{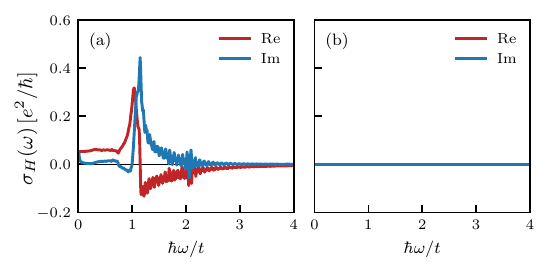}
    \caption{Real (red) and imaginary (blue) part of the anomalous Hall conductivity $\sigma_H$ as a function of optical frequency $\hbar \omega$ normalized with $t$ for chiral $d$-wave spin-singlet pairing on the honeycomb lattice with Kane-Mele SOC (a) with interband pairing and (b) without interband pairing. Here $\lambda_{\rm SOC}=0.05 t$, $\mu=-0.5 t$, $\Delta_0=0.1 t$, and temperature $k_B T= 0.05t$. }
    \label{fig:Fig1}
\end{figure}
%%%%%%%%%%%%%%%%%%%%%%%%%%%%%%%%%%%%%%%%%%%%%%%%%%%%%%%%%%%%%%%%%%%%%%%%%%%%%%%%%%%%%%%%%%%%%%%%%%%%%%%%%%%%%%%%%%%%%%%%%%%%%%%%%%%%%%

\noindent
\textcolor{blue}{\textit{Intrinsic annomlaous Hall conductivity.}} Having establish the criteria for nonunitarity, we next calculate the Hall conductivity of a superconductor at optical frequencies, $\sigma_{H}(\omega)$, using Kubo linear response theory. Up to first loop order~\cite{Brydon21,Zhang24}
\begin{align}
    \sigma_H (\omega) = \frac{i}{2\omega}\lim_{i\omega_n \to \omega + i0^+}[ \pi_{yx} (i\omega_n )- \pi_{xy} (i\omega_n) ],\label{eq:hall conductivity kuba formula1}
\end{align}
where
\begin{small}
    \begin{align}
    \pi_{ab} ( i \omega_n) = \frac{e^2}{2 N} \sum_{\mathbf{k}} \frac{1}{\beta} \sum_{i \nu_m} \mathrm{Tr}[\mathcal{V}^a_{\mk} \mathcal{G}_\mk(i\omega_n+i\nu_m)\mathcal{V}^b_{\mk}\mathcal{G}_\mk(i\nu_m) ],\label{eq:hall conductivity kuba formula2}
\end{align}
\end{small}
\noindent
in which $\nu_m = (2m+1)\pi/\beta$ and $\omega_n = 2n\pi/\beta$ are the fermioinic and bosonic Matsubara frequencies respectively, and $\mathcal{G}_{\mathbf{k}, i\nu_m}= [i \nu_m- \mathcal{H}_{\text{BdG}}(\mathbf{k})]^{-1}$ is the Gorkov Green's function of the corresponding BdG Hamiltonian in Eq~(\ref{eq:BdG Hamiltonian}). Further, $e$ is the electronic charge, $\beta = 1/k_BT$ the inverse temperature, and $N$ is the number of sampling points in the Brillouin zone. 
The velocity operator matrix in the Nambu basis reads as
\begin{align}
    \mathcal{V}^a_\mathbf{k} = 
    \begin{pmatrix}
        V_{a,\mathbf{k}} & 0 \\
        0 & -(V_{a,\bar{\mathbf{k}}})^T
    \end{pmatrix},
\end{align}
where the matrix $  V_{a,\mathbf{k}}=\partial_{k_a} H_N (\mathbf{k})$. In addition to broken TRS, it follows from Eqs.~(\ref{eq:hall conductivity kuba formula1}-\ref{eq:hall conductivity kuba formula2}) that a necessary condition for a finite Hall response is that the Green’s function and the velocity matrix do not commute, i.e., $ [\mathcal{G}_\mathbf{k}, \mathcal{V}^a_\mathbf{k} ] \neq 0$.  This condition can be satisfied when the system exhibits finite interband pairing, since the Hamiltonian and consequently the Green’s function then contains terms that couple distinct bands having different group velocities. However, in this work we focus on effects only from intraband pairing. Then, in the absence of interband pairing, the Green’s function becomes block diagonal and a nonvanishing commutator instead relies on the presence of an interband velocity term that couples different bands. The velocity matrix in the band basis, i.e. where $H_N(\mathbf{k})$ is diagonal, takes the form
\begin{align}
    \tilde{V}^{mn}_{a,\mathbf{k}} = \delta_{mn}\partial_{k_a}\epsilon_{m\mathbf{k}}  + (\epsilon_{m\mathbf{k}} - \epsilon_{n\mathbf{k}})\langle{\partial_{k_a}\psi_{m,\mathbf{k}} }\ket{\psi_{n,\mathbf{k}}},
    \label{eq:velocity_matrix_band_basis}
\end{align}
where $\psi_{m,\mathbf{k}}$ is the eigenstate of $H_N(\mathbf{k})$ with energy $\epsilon_{m\mathbf{k}}$ and we use the tilde notation to denote the band basis for the velocity matrix. 
The first term corresponds to the usual group velocity, while the second term accounts for the interband velocity $\langle{\partial_{k_a}\psi_{m,\mathbf{k}} }\ket{\psi_{n,\mathbf{k}}}$, which encodes the quantum geometric properties of the Bloch bands~\cite{Torma23}. 
A finite interband velocity indicates coherent transport of electrons between distinct Bloch bands.

If we assume purely intraband pairing and restrict to either equal- or opposite-spin channels, Eq.~(\ref{eq:hall conductivity kuba formula1}) simplifies to 
\begin{align}
    \sigma_H(\omega) = \sum_{s=\uparrow,\downarrow} \sigma_H^{s,\zeta(s)},
\end{align}
with
\begin{align}
     \sigma^{s, \zeta(s)}_H(\omega) 
     = \frac{ie^2}{4N} \sum_{\mk} &\frac{\epsilon_{\mk1}/E_{\mk 1 s \zeta(s)} - \epsilon_{\mk 2}/E_{\mk 2 s \zeta(s)}}{(\omega + i0^+)^2 - (E_{\mk 1 s \zeta(s)} + E_{\mk 2 s \zeta(s)})^2} \nonumber \\
     &\times (Q^{e, s}_{xy} -Q^{h, \zeta(s)}_{xy} ), \label{eq:hall conductivity simplified both cases}
\end{align}
where $\zeta(s) = s$ or $\zeta(s) = \bar{s}$ for pairing between equal or opposite spin, respectively, and $E_{\mk n s \zeta(s)}$ are the eigenvalues of the BdG Hamiltonian, enumerated by $n$ and spin indices $s$ and $\zeta(s)$.
%= (\epsilon^2_{n\mathbf{k}} + |\Delta^{s,\eta(s)}_{n\mathbf{k}}|^2)^{1/2}
Here, $Q$ originates from virtual interband transitions and can be expressed in terms of the velocity matrix elements as 
\begin{align}
    Q^{e, s}_{xy} &= \tilde{V}_{x\mk s}^{12}\tilde{V}_{y\mk s}^{21} - \tilde{V}_{y\mk s}^{12}\tilde{V}_{x\mk s}^{21}, \\
    Q^{h, s}_{xy} &= (\tilde{V}_{x\bar{\mk} s}^T)^{12}(\tilde{V}_{y\bar{\mk} s}^T)^{21} - (\tilde{V}_{y\bar{\mk} s}^T)^{12}(\tilde{V}_{x\bar{\mk} s}^T)^{21}.
\end{align}
The $Q$'s are related to the normal-state spin-Berry curvature $\Omega^{s}_{n,xy}$ as
\begin{align}
    Q^{e, s}_{xy} & = -i(\epsilon_{\mathbf{k}1 s} - \epsilon_{\mathbf{k}2 s})^2\,\Omega^s_{1,xy}(\mathbf{k}),\\ 
    Q^{h, \zeta(s)}_{xy} & = i(\epsilon_{\bar{\mathbf{k}} 1 \zeta(s)} - \epsilon_{\bar{\mathbf{k}} 2 \zeta(s)})^2\,\Omega^{\zeta(s)}_{1,xy}(\bar{\mathbf{k}}).
    %\text{Check it!}
\end{align}
Details of the derivation for both pairing cases are presented in the SM~\cite{SM}.

In arriving at the unified equation of the Hall conductivity in Eq.~(\ref{eq:hall conductivity simplified both cases}), we have utilized the presence of both inversion and TRS in the normal-state Hamiltonian $H_N (\mathbf{k})$, such that in the normal state $\epsilon_{i\bar{\mk},\downarrow} = \epsilon_{i\mk,\uparrow}=\epsilon_{i\mk,\downarrow}=\epsilon_{i\mk}$.
In the case of spin-singlet pairing, $Q^{e, s}_{xy}= Q^{h, \zeta(s)}_{xy}$, since the Berry curvature of the TRS preserving normal-state obeys $\Omega^\uparrow_{1,xy}(\mathbf{k}) = -\Omega^\downarrow_{1,xy}(-\mathbf{k})$, leading to $\sigma_H (\omega)=0$ in Eq.~(\ref{eq:hall conductivity simplified both cases}). In contrast, the $Q$'s add up for the case of equal spin-triplet pairing, which may result in nonvanishing $\sigma_H (\omega)$. This follows from the presence of inversion symmetry in the normal state, resulting in $\Omega^\uparrow_{1,xy}(\mathbf{k}) = \Omega^\uparrow_{1,xy}(-\mathbf{k})$. For spin-triplet pairing, $\sigma_H (\omega)$ then simplifies to
{\begin{small}
\begin{align}
\sigma_H(\omega)
&= \frac{ie^2}{2N} \sum_{\mk} (\epsilon_{\mk 1} - \epsilon_{\mk 2})^2\,\Omega^\uparrow_{1,xy}(\mathbf{k}) \nonumber \\
&\times \Bigg[
\frac{\epsilon_{1\mk}/E_{\mk 1 \uparrow \uparrow} - \epsilon_{2\mk}/E_{\mk 2\uparrow \uparrow}}{D_{\uparrow \uparrow}}
-
\frac{\epsilon_{\mk 1}/E_{\mk 1 \downarrow \downarrow} - \epsilon_{\mk 2}/E_{\mk 2 \downarrow \downarrow}}{D_{\downarrow \downarrow}}
\Bigg],
\end{align}
\end{small}}
where $D_{ss} = (\omega + i0^+)^2 - (E_{\mk 1 ss}+E_{\mk 2 ss})^2$. This establishes that a finite $\sigma_H (\omega)$ arises even in the absence of interband pairing, provided that the normal-state exhibits finite a spin-Hall effect, i.e., $\Omega^\uparrow_{1,xy}(\mathbf{k})\neq 0$ and, additionally, the superconducting state is spin-polarized, i.e., $E_{\mathbf{k}i \uparrow\uparrow} \neq E_{\mathbf{k} i \downarrow\downarrow}$.
To understand these results, we note that in the spin-singlet case, $Q^{e,12}_{xy} - Q^{h,12}_{xy} = 0$ forces $\sigma_H (\omega)$ to vanish, independently of the BdG spectra.
This cancellation originates from the nature of the pairing. In the spin-singlet channel, Cooper pairs form between time-reversed partners $(\mathbf{k}\uparrow, -\mathbf{k}\downarrow)$, which, due to the presence of TRS in the normal state, carry opposite Berry curvature, $\Omega_1^\uparrow(\mathbf{k}) = -\Omega_1^\downarrow(-\mathbf{k})$.
In contrast, for equal-spin triplet pairing, pairing occurs between $(\mathbf{k}\uparrow, -\mathbf{k}\uparrow)$, which share the same Berry curvature in the presence of inversion symmetry in the normal state and therefore do not exhibit the same cancellation. 

Importantly, thus far, we have not specified the nature of the equal-spin-triplet pairing state, namely whether it is  unitary or nonunitary in order to exhibit a finite AHR. However, the results above show that a finite AHR in addition requires the superconducting state to be spin-polarized.
The possibility of a spin-polarized superconducting state has traditionally been associated with nonunitary spin-triplet pairing, seemingly ruling out the possibility of a finite AHR in unitary spin-triplet states. Next we will show that the necessary spin-polarization is also achievable in spin-unitary superconductors.\\
\begin{figure}[t!]
    \centering
    \includegraphics[width=\columnwidth]{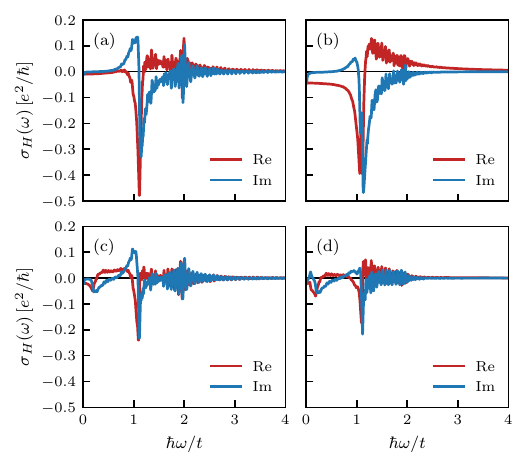}
    \caption{Real (red) and imaginary (blue) part of the anomalous Hall conductivity $\sigma_H$ as a function of optical frequency $\hbar \omega$ normalized with $t$ for chiral $p$-wave equal-spin triplet pairing on the honeycomb lattice with Kane-Mele SOC for 
    (a) interband pairing and $\kappa=0$, 
    (b) interband pairing and $\kappa=1$, 
    (c) no interband pairing and $\kappa=0$, and
    (d) no interband pairing and $\kappa=1$. Other parameters are the same as in Fig.~\ref{fig:Fig1}.}
    \label{fig:Fig2}
\end{figure}
\noindent
\textcolor{blue}{\textit{Spin polarization from nonunitary orbital pairing.} }To demonstrate that a spin-unitary state can possess a spin-polarized superconducting state, and thereby give a finite AHR, we start by considering a general normal-state Hamiltonian written in a spin-resolved basis, such that the spin sectors are decoupled ($S_z$ is a good quantum number)
\begin{align}
    \mathcal{H}_{\text{normal}}(\mathbf{k})=\begin{pmatrix}
        H^\uparrow_N(\mathbf{k}) & 0 \\
        0 & H^\downarrow_N(\mathbf{k})
    \end{pmatrix},
\end{align}
where $H_N^s(\mathbf{k}) = a_0 \tau_0 + \vec{a}_s\cdot \vec{\tau}$, such that $\vec{a}_s=\{a_1,a_2,s a_3\}$ and $s=+1$ for $\uparrow$ and $s=-1$ for $\downarrow$. We note that $\mathcal{H}_\text{normal}(\mathbf{k})$ includes SOC that preserves $S_z$, and thus includes also the Hamiltonian in Eq.~(\ref{eq:normal state Hamiltonian}). 
We then consider an equal-spin triplet spin-unitary pairing state, such that $\Delta_{\text{spin}} = \sigma_0$, owing to $\kappa=1$, see Eq~(\ref{eq:delta_spin}). We finally introduce generic nonunitary pairing matrix in orbital space of the form $\Delta_{\text{orb}}= b_1 \, \tau_1 + i b_2 \, \tau_2$, which is consistent with the form considered in Eq.~(\ref{eq:orbital_pairing_matrix}). Consequently, the criterion to determine nonunitarity is 
$\Delta \Delta^\dagger  = \sigma_0 \otimes (u \tau_0 + v \tau_3)$, where $u = b_1^2 + b_2^2$ and $v= 2 b_1 b_2$. A finite $v$ signals nonunitarity in the orbital sector, similar to Eq.~(\ref{eq: Deltao_nonunitary}). 

For the equal-spin triplet pairing, the spin-up and spin-down sectors remain decoupled, allowing the full $8\times8$ (Nambu $\otimes$ spin $\otimes$ orbital) BdG Hamiltonian to be block diagonalized into two independent $4\times4$  BdG Hamiltonians $H_{\text{BdG}} = H^\uparrow_{\text{BdG},4\times4} \oplus H^\downarrow_{\text{BdG},4\times4}$. Each block inherits particle–hole symmetry, implying a symmetric spectrum $\{\pm E_{s,+},\pm E_{s,-}\}$. Consequently, the eigenvalues of the reduced BdG Hamiltonian can be obtained by solving the characteristic equation, which reduces to a quartic form,
\begin{align}
    E^4 + p E^2 + q =0,
\end{align}
with $p = -\frac{1}{2}\,\mathrm{Tr}\!\left[\left(H_{\mathrm{BdG},\,4\times4}^{s}\right)^2\right]$ and $q = \det[H^s_{\text{BdG},4\times4}]$. The four eigenvalues are then given by
\begin{align}
    E^2_{\pm}= \frac{-p  \pm \sqrt{p^2-4q}}{2}.
\end{align}
Substituting the explicit form of $H^s_{\text{BdG}}$, the quasiparticle spectrum can be written as
\begin{equation}
\resizebox{\columnwidth}{!}{$
E^2_{s,\pm}= a_0^2 + |\vec{a_s}|^2 + u \pm 2 \sqrt{a_0^2 |\vec{a_s}|^2+ a_2^2 b_1^2 + (a_1^2 +b_1^2) b_2^2 + s a_0 a_3 v}.
$}
\label{eq:Espin}
\end{equation}
Crucially, the BdG spectrum becomes spin-dependent only if $ v \neq 0$ and in the presence of the spin-dependent $a_3$ term. This establishes that even a spin-unitary pairing matrix can give rise to spin-polarized BdG spectra, when the orbital-space pairing matrix is nonunitary and SOC is present.

Finally, we apply the results of Eq.~(\ref{eq:Espin}) to the honeycomb lattice with equal-spin triplet and spin-unitary pairing. Nonzero orbital nonunitarity, i.e.~nonzero $q_{\pm}$ in Eq.~(\ref{eq:orbital_nonunitary_def}) thus yields a spin-split BdG spectra in the presence of Kane-Mele SOC $\lambda_{\rm SOC}$. The presence of finite $q_{\pm}$ signifies a nonzero induced sublattice polarization, $ \mathrm{Tr}[\Delta_{\mathbf{k}}^\dagger \sigma_0\otimes\tau_3 \Delta_{\mathbf{k}}] / 4$ in the pairing state, which signifies a breaking of the sublattice symmetry by the superconducting order parameter. It is this sublattice polarization, which in the presence of the Kane-Mele SOC, can be viewed to act as the mechanism that lifts the spin degeneracy of the BdG spectra. As a consequence, even a spin-unitary state can provide the spin-polarized condensate necessary to obtain a finite AHR even for pure intraband pairing. 

\noindent
\textcolor{blue}{\textit{Numerical Results.}} We finally turn to numerical evaluation of the AHR in Eq.~(\ref{eq:hall conductivity kuba formula1}) using Eq.~(\ref{eq:hall conductivity kuba formula2}).  For chiral $d$-wave spin-singlet pairing in a TRS preserving normal state, AHR is finite when including interband pairing, see Fig.~\ref{fig:Fig1}(a), while removing the interband pairing by hand results in a vanishing AHR, see Fig.~\ref{fig:Fig1}(b). This is in full agreement with our analytical findings.
Then, we consider the equal-spin triplet case in Fig.~\ref{fig:Fig2} for both the spin-unitary and spin-nonunitary cases, both with and without interband pairing. In Figs.~\ref{fig:Fig2}(a,c), we set $\kappa = \Delta_{\downarrow\downarrow}/\Delta_{\uparrow\uparrow} = 0$, corresponding to a spin-nonunitary pairing. The interband pairing is finite in Fig.~\ref{fig:Fig2}(a) and vanishes in Fig.~\ref{fig:Fig2}(c). In both cases, the AHR remains finite. In Figs.~\ref{fig:Fig2}(b,d), we consider $\kappa = 1$, corresponding to a spin-unitary pairing. The interband pairing is again finite in Fig.~\ref{fig:Fig2}(b) and absent in Fig.~\ref{fig:Fig2}(d). In both cases, the AHR remains finite. Of particular interest is Fig.~\ref{fig:Fig2}(d), where the Hall response persists even for spin-unitary pairing, even in the absence of interband pairing. Based on our analytical derivation, the results in Fig.~\ref{fig:Fig2}(d) can be explained by a finite spin-polarization of the BdG spectrum induced by orbital nonunitarity and Kane-Mele SOC, together with the presence of a nontrivial quantum geometry encoded in the finite spin Berry curvature.
\noindent
%%%%%%%%%%%%%%%%%%%%%%%%%%%%%%%%%%%%%%%%%%%%%%%%%%%%%%%%%%%%%%%%%%%%%%%%%%%%%%%%%%%%%%%%%%%%%%%%%%%%%%%%%%%%%%%%%%%%%%%%%%%%%%%%%%%%%%%%%%%%%%%%%%%%%%%%%%%%%%%%%%%%%%%%%%%%%%%%%%%%%%%%%%%%%%%%%%%%%%%%%%%%%%%%%%%%%%

\textcolor{blue}{\textit{Conclusions.}} In summary, we investigate the conditions under which a finite AHR can arise in multiband superconductors in the absence of interband pairing, driven instead by nontrivial orbital pairing structure and quantum geometric effects. Using a honeycomb lattice model with Kane–Mele SOC, we show that a finite AHR can emerge from pure intraband pairing through the interplay of band geometry and orbital nonunitarity, the latter providing the necessary spin splitting of the BdG spectrum even for a spin-unitary state when SOC is present. In contrast to earlier studies~\cite{Brydon19} in the same model that required interband pairing and neglected Kane–Mele SOC, our results demonstrate that an equal-spin triplet state can support a finite AHR, even while remaining unitary in the spin sector due to a  nontrivial orbital (sublattice) structure. However, a chiral spin-singlet state, also with a nonunitary orbital structure, does not exhibit AHR. 
Our results thus challenge the conventional view that spin nonunitarity or interband pairing is necessary for a finite Kerr signal and instead highlight the broader role of quantum geometry and multiorbital effects in determining electromagnetic responses in superconductors. More generally, our results suggest that experimentally observed Kerr signals, which are directly related to a finite AHR, may originate from a broader class of pairing states than previously anticipated.

\noindent
{\textcolor{blue}{\textit{Acknowledgement.}}} We acknowledge financial support from the Knut and Alice Wallenberg Foundation through the Wallenberg Scholar program, KAW 2023.0244 and the European Union through the European Research Council (ERC) under the European Union’s Horizon 2020 research and innovation programme (ERC-2022-CoG, Grant agreement No.~101087096). Views and opinions expressed are however those of the author(s) only and do not necessarily reflect those of the European Union or the European Research Council Executive Agency. Neither the European Union nor the granting authority can be held responsible for them.
\bibliography{reference}

\begin{thebibliography}{35}%
\makeatletter
\providecommand \@ifxundefined [1]{%
 \@ifx{#1\undefined}
}%
\providecommand \@ifnum [1]{%
 \ifnum #1\expandafter \@firstoftwo
 \else \expandafter \@secondoftwo
 \fi
}%
\providecommand \@ifx [1]{%
 \ifx #1\expandafter \@firstoftwo
 \else \expandafter \@secondoftwo
 \fi
}%
\providecommand \natexlab [1]{#1}%
\providecommand \enquote  [1]{``#1''}%
\providecommand \bibnamefont  [1]{#1}%
\providecommand \bibfnamefont [1]{#1}%
\providecommand \citenamefont [1]{#1}%
\providecommand \href@noop [0]{\@secondoftwo}%
\providecommand \href [0]{\begingroup \@sanitize@url \@href}%
\providecommand \@href[1]{\@@startlink{#1}\@@href}%
\providecommand \@@href[1]{\endgroup#1\@@endlink}%
\providecommand \@sanitize@url [0]{\catcode `\\12\catcode `\$12\catcode `\&12\catcode `\#12\catcode `\^12\catcode `\_12\catcode `\%12\relax}%
\providecommand \@@startlink[1]{}%
\providecommand \@@endlink[0]{}%
\providecommand \url  [0]{\begingroup\@sanitize@url \@url }%
\providecommand \@url [1]{\endgroup\@href {#1}{\urlprefix }}%
\providecommand \urlprefix  [0]{URL }%
\providecommand \Eprint [0]{\href }%
\providecommand \doibase [0]{https://doi.org/}%
\providecommand \selectlanguage [0]{\@gobble}%
\providecommand \bibinfo  [0]{\@secondoftwo}%
\providecommand \bibfield  [0]{\@secondoftwo}%
\providecommand \translation [1]{[#1]}%
\providecommand \BibitemOpen [0]{}%
\providecommand \bibitemStop [0]{}%
\providecommand \bibitemNoStop [0]{.\EOS\space}%
\providecommand \EOS [0]{\spacefactor3000\relax}%
\providecommand \BibitemShut  [1]{\csname bibitem#1\endcsname}%
\let\auto@bib@innerbib\@empty
%</preamble>
\bibitem [{\citenamefont {Xia}\ \emph {et~al.}(2006)\citenamefont {Xia}, \citenamefont {Maeno}, \citenamefont {Beyersdorf}, \citenamefont {Fejer},\ and\ \citenamefont {Kapitulnik}}]{Kapitunik06}%
  \BibitemOpen
  \bibfield  {author} {\bibinfo {author} {\bibfnamefont {J.}~\bibnamefont {Xia}}, \bibinfo {author} {\bibfnamefont {Y.}~\bibnamefont {Maeno}}, \bibinfo {author} {\bibfnamefont {P.~T.}\ \bibnamefont {Beyersdorf}}, \bibinfo {author} {\bibfnamefont {M.~M.}\ \bibnamefont {Fejer}},\ and\ \bibinfo {author} {\bibfnamefont {A.}~\bibnamefont {Kapitulnik}},\ }\bibfield  {title} {\bibinfo {title} {High resolution polar kerr effect measurements of ${\mathrm{sr}}_{2}{\mathrm{ruo}}_{4}$: Evidence for broken time-reversal symmetry in the superconducting state},\ }\href {https://doi.org/10.1103/PhysRevLett.97.167002} {\bibfield  {journal} {\bibinfo  {journal} {Phys. Rev. Lett.}\ }\textbf {\bibinfo {volume} {97}},\ \bibinfo {pages} {167002} (\bibinfo {year} {2006})}\BibitemShut {NoStop}%
\bibitem [{\citenamefont {Hayes}\ \emph {et~al.}(2021)\citenamefont {Hayes}, \citenamefont {Wei}, \citenamefont {Metz}, \citenamefont {Zhang}, \citenamefont {Eo}, \citenamefont {Ran}, \citenamefont {Saha}, \citenamefont {Collini}, \citenamefont {Butch}, \citenamefont {Agterberg} \emph {et~al.}}]{Hayes21}%
  \BibitemOpen
  \bibfield  {author} {\bibinfo {author} {\bibfnamefont {I.~M.}\ \bibnamefont {Hayes}}, \bibinfo {author} {\bibfnamefont {D.~S.}\ \bibnamefont {Wei}}, \bibinfo {author} {\bibfnamefont {T.}~\bibnamefont {Metz}}, \bibinfo {author} {\bibfnamefont {J.}~\bibnamefont {Zhang}}, \bibinfo {author} {\bibfnamefont {Y.~S.}\ \bibnamefont {Eo}}, \bibinfo {author} {\bibfnamefont {S.}~\bibnamefont {Ran}}, \bibinfo {author} {\bibfnamefont {S.~R.}\ \bibnamefont {Saha}}, \bibinfo {author} {\bibfnamefont {J.}~\bibnamefont {Collini}}, \bibinfo {author} {\bibfnamefont {N.~P.}\ \bibnamefont {Butch}}, \bibinfo {author} {\bibfnamefont {D.~F.}\ \bibnamefont {Agterberg}}, \emph {et~al.},\ }\bibfield  {title} {\bibinfo {title} {Multicomponent superconducting order parameter in ute2},\ }\href@noop {} {\bibfield  {journal} {\bibinfo  {journal} {Science}\ }\textbf {\bibinfo {volume} {373}},\ \bibinfo {pages} {797} (\bibinfo {year} {2021})}\BibitemShut {NoStop}%
\bibitem [{\citenamefont {Schemm}\ \emph {et~al.}(2014)\citenamefont {Schemm}, \citenamefont {Gannon}, \citenamefont {Wishne}, \citenamefont {Halperin},\ and\ \citenamefont {Kapitulnik}}]{Schemm14}%
  \BibitemOpen
  \bibfield  {author} {\bibinfo {author} {\bibfnamefont {E.}~\bibnamefont {Schemm}}, \bibinfo {author} {\bibfnamefont {W.}~\bibnamefont {Gannon}}, \bibinfo {author} {\bibfnamefont {C.}~\bibnamefont {Wishne}}, \bibinfo {author} {\bibfnamefont {W.}~\bibnamefont {Halperin}},\ and\ \bibinfo {author} {\bibfnamefont {A.}~\bibnamefont {Kapitulnik}},\ }\bibfield  {title} {\bibinfo {title} {Observation of broken time-reversal symmetry in the heavy-fermion superconductor upt3},\ }\href@noop {} {\bibfield  {journal} {\bibinfo  {journal} {Science}\ }\textbf {\bibinfo {volume} {345}},\ \bibinfo {pages} {190} (\bibinfo {year} {2014})}\BibitemShut {NoStop}%
\bibitem [{\citenamefont {Kapitulnik}\ \emph {et~al.}(2009)\citenamefont {Kapitulnik}, \citenamefont {Xia}, \citenamefont {Schemm},\ and\ \citenamefont {Palevski}}]{Kapitulnik_2009}%
  \BibitemOpen
  \bibfield  {author} {\bibinfo {author} {\bibfnamefont {A.}~\bibnamefont {Kapitulnik}}, \bibinfo {author} {\bibfnamefont {J.}~\bibnamefont {Xia}}, \bibinfo {author} {\bibfnamefont {E.}~\bibnamefont {Schemm}},\ and\ \bibinfo {author} {\bibfnamefont {A.}~\bibnamefont {Palevski}},\ }\bibfield  {title} {\bibinfo {title} {Polar kerr effect as probe for time-reversal symmetry breaking in unconventional superconductors},\ }\href {https://doi.org/10.1088/1367-2630/11/5/055060} {\bibfield  {journal} {\bibinfo  {journal} {New J. Phys.}\ }\textbf {\bibinfo {volume} {11}},\ \bibinfo {pages} {055060} (\bibinfo {year} {2009})}\BibitemShut {NoStop}%
\bibitem [{\citenamefont {Nagaosa}\ \emph {et~al.}(2010)\citenamefont {Nagaosa}, \citenamefont {Sinova}, \citenamefont {Onoda}, \citenamefont {MacDonald},\ and\ \citenamefont {Ong}}]{Nagaosa10}%
  \BibitemOpen
  \bibfield  {author} {\bibinfo {author} {\bibfnamefont {N.}~\bibnamefont {Nagaosa}}, \bibinfo {author} {\bibfnamefont {J.}~\bibnamefont {Sinova}}, \bibinfo {author} {\bibfnamefont {S.}~\bibnamefont {Onoda}}, \bibinfo {author} {\bibfnamefont {A.~H.}\ \bibnamefont {MacDonald}},\ and\ \bibinfo {author} {\bibfnamefont {N.~P.}\ \bibnamefont {Ong}},\ }\bibfield  {title} {\bibinfo {title} {Anomalous hall effect},\ }\href {https://doi.org/10.1103/RevModPhys.82.1539} {\bibfield  {journal} {\bibinfo  {journal} {Rev. Mod. Phys.}\ }\textbf {\bibinfo {volume} {82}},\ \bibinfo {pages} {1539} (\bibinfo {year} {2010})}\BibitemShut {NoStop}%
\bibitem [{\citenamefont {Xiao}\ \emph {et~al.}(2010)\citenamefont {Xiao}, \citenamefont {Chang},\ and\ \citenamefont {Niu}}]{Xiao10}%
  \BibitemOpen
  \bibfield  {author} {\bibinfo {author} {\bibfnamefont {D.}~\bibnamefont {Xiao}}, \bibinfo {author} {\bibfnamefont {M.-C.}\ \bibnamefont {Chang}},\ and\ \bibinfo {author} {\bibfnamefont {Q.}~\bibnamefont {Niu}},\ }\bibfield  {title} {\bibinfo {title} {Berry phase effects on electronic properties},\ }\href {https://doi.org/10.1103/RevModPhys.82.1959} {\bibfield  {journal} {\bibinfo  {journal} {Rev. Mod. Phys.}\ }\textbf {\bibinfo {volume} {82}},\ \bibinfo {pages} {1959} (\bibinfo {year} {2010})}\BibitemShut {NoStop}%
\bibitem [{\citenamefont {Roy}\ and\ \citenamefont {Kallin}(2008)}]{Roy08}%
  \BibitemOpen
  \bibfield  {author} {\bibinfo {author} {\bibfnamefont {R.}~\bibnamefont {Roy}}\ and\ \bibinfo {author} {\bibfnamefont {C.}~\bibnamefont {Kallin}},\ }\bibfield  {title} {\bibinfo {title} {Collective modes and electromagnetic response of a chiral superconductor},\ }\href {https://doi.org/10.1103/PhysRevB.77.174513} {\bibfield  {journal} {\bibinfo  {journal} {Phys. Rev. B}\ }\textbf {\bibinfo {volume} {77}},\ \bibinfo {pages} {174513} (\bibinfo {year} {2008})}\BibitemShut {NoStop}%
\bibitem [{\citenamefont {Lutchyn}\ \emph {et~al.}(2008)\citenamefont {Lutchyn}, \citenamefont {Nagornykh},\ and\ \citenamefont {Yakovenko}}]{Lutchyn08}%
  \BibitemOpen
  \bibfield  {author} {\bibinfo {author} {\bibfnamefont {R.~M.}\ \bibnamefont {Lutchyn}}, \bibinfo {author} {\bibfnamefont {P.}~\bibnamefont {Nagornykh}},\ and\ \bibinfo {author} {\bibfnamefont {V.~M.}\ \bibnamefont {Yakovenko}},\ }\bibfield  {title} {\bibinfo {title} {Gauge-invariant electromagnetic response of a chiral ${p}_{x}+i{p}_{y}$ superconductor},\ }\href {https://doi.org/10.1103/PhysRevB.77.144516} {\bibfield  {journal} {\bibinfo  {journal} {Phys. Rev. B}\ }\textbf {\bibinfo {volume} {77}},\ \bibinfo {pages} {144516} (\bibinfo {year} {2008})}\BibitemShut {NoStop}%
\bibitem [{\citenamefont {Read}\ and\ \citenamefont {Green}(2000)}]{Read2000}%
  \BibitemOpen
  \bibfield  {author} {\bibinfo {author} {\bibfnamefont {N.}~\bibnamefont {Read}}\ and\ \bibinfo {author} {\bibfnamefont {D.}~\bibnamefont {Green}},\ }\bibfield  {title} {\bibinfo {title} {Paired states of fermions in two dimensions with breaking of parity and time-reversal symmetries and the fractional quantum hall effect},\ }\href {https://doi.org/10.1103/PhysRevB.61.10267} {\bibfield  {journal} {\bibinfo  {journal} {Phys. Rev. B}\ }\textbf {\bibinfo {volume} {61}},\ \bibinfo {pages} {10267} (\bibinfo {year} {2000})}\BibitemShut {NoStop}%
\bibitem [{\citenamefont {Goryo}(2008)}]{Goryo08}%
  \BibitemOpen
  \bibfield  {author} {\bibinfo {author} {\bibfnamefont {J.}~\bibnamefont {Goryo}},\ }\bibfield  {title} {\bibinfo {title} {Impurity-induced polar kerr effect in a chiral $p$-wave superconductor},\ }\href {https://doi.org/10.1103/PhysRevB.78.060501} {\bibfield  {journal} {\bibinfo  {journal} {Phys. Rev. B}\ }\textbf {\bibinfo {volume} {78}},\ \bibinfo {pages} {060501} (\bibinfo {year} {2008})}\BibitemShut {NoStop}%
\bibitem [{\citenamefont {Lutchyn}\ \emph {et~al.}(2009)\citenamefont {Lutchyn}, \citenamefont {Nagornykh},\ and\ \citenamefont {Yakovenko}}]{Lutchyn09}%
  \BibitemOpen
  \bibfield  {author} {\bibinfo {author} {\bibfnamefont {R.~M.}\ \bibnamefont {Lutchyn}}, \bibinfo {author} {\bibfnamefont {P.}~\bibnamefont {Nagornykh}},\ and\ \bibinfo {author} {\bibfnamefont {V.~M.}\ \bibnamefont {Yakovenko}},\ }\bibfield  {title} {\bibinfo {title} {Frequency and temperature dependence of the anomalous ac hall conductivity in a chiral ${p}_{x}+i{p}_{y}$ superconductor with impurities},\ }\href {https://doi.org/10.1103/PhysRevB.80.104508} {\bibfield  {journal} {\bibinfo  {journal} {Phys. Rev. B}\ }\textbf {\bibinfo {volume} {80}},\ \bibinfo {pages} {104508} (\bibinfo {year} {2009})}\BibitemShut {NoStop}%
\bibitem [{\citenamefont {K\"onig}\ and\ \citenamefont {Levchenko}(2017)}]{Levchenko17}%
  \BibitemOpen
  \bibfield  {author} {\bibinfo {author} {\bibfnamefont {E.~J.}\ \bibnamefont {K\"onig}}\ and\ \bibinfo {author} {\bibfnamefont {A.}~\bibnamefont {Levchenko}},\ }\bibfield  {title} {\bibinfo {title} {Kerr effect from diffractive skew scattering in chiral ${p}_{x}\ifmmode\pm\else\textpm\fi{}i{p}_{y}$ superconductors},\ }\href {https://doi.org/10.1103/PhysRevLett.118.027001} {\bibfield  {journal} {\bibinfo  {journal} {Phys. Rev. Lett.}\ }\textbf {\bibinfo {volume} {118}},\ \bibinfo {pages} {027001} (\bibinfo {year} {2017})}\BibitemShut {NoStop}%
\bibitem [{\citenamefont {Li}\ \emph {et~al.}(2020)\citenamefont {Li}, \citenamefont {Wang},\ and\ \citenamefont {Huang}}]{Li20}%
  \BibitemOpen
  \bibfield  {author} {\bibinfo {author} {\bibfnamefont {Y.}~\bibnamefont {Li}}, \bibinfo {author} {\bibfnamefont {Z.}~\bibnamefont {Wang}},\ and\ \bibinfo {author} {\bibfnamefont {W.}~\bibnamefont {Huang}},\ }\bibfield  {title} {\bibinfo {title} {Anomalous hall effect in single-band chiral superconductors from impurity superlattices},\ }\href {https://doi.org/10.1103/PhysRevResearch.2.042027} {\bibfield  {journal} {\bibinfo  {journal} {Phys. Rev. Res.}\ }\textbf {\bibinfo {volume} {2}},\ \bibinfo {pages} {042027} (\bibinfo {year} {2020})}\BibitemShut {NoStop}%
\bibitem [{\citenamefont {Liu}\ \emph {et~al.}(2023)\citenamefont {Liu}, \citenamefont {Chen},\ and\ \citenamefont {Huang}}]{Liu23}%
  \BibitemOpen
  \bibfield  {author} {\bibinfo {author} {\bibfnamefont {H.-T.}\ \bibnamefont {Liu}}, \bibinfo {author} {\bibfnamefont {W.}~\bibnamefont {Chen}},\ and\ \bibinfo {author} {\bibfnamefont {W.}~\bibnamefont {Huang}},\ }\bibfield  {title} {\bibinfo {title} {Impact of random impurities on the anomalous hall effect in chiral superconductors},\ }\href {https://doi.org/10.1103/PhysRevB.107.224517} {\bibfield  {journal} {\bibinfo  {journal} {Phys. Rev. B}\ }\textbf {\bibinfo {volume} {107}},\ \bibinfo {pages} {224517} (\bibinfo {year} {2023})}\BibitemShut {NoStop}%
\bibitem [{\citenamefont {Taylor}\ and\ \citenamefont {Kallin}(2012)}]{kallin12}%
  \BibitemOpen
  \bibfield  {author} {\bibinfo {author} {\bibfnamefont {E.}~\bibnamefont {Taylor}}\ and\ \bibinfo {author} {\bibfnamefont {C.}~\bibnamefont {Kallin}},\ }\bibfield  {title} {\bibinfo {title} {Intrinsic hall effect in a multiband chiral superconductor in the absence of an external magnetic field},\ }\href {https://doi.org/10.1103/PhysRevLett.108.157001} {\bibfield  {journal} {\bibinfo  {journal} {Phys. Rev. Lett.}\ }\textbf {\bibinfo {volume} {108}},\ \bibinfo {pages} {157001} (\bibinfo {year} {2012})}\BibitemShut {NoStop}%
\bibitem [{\citenamefont {P.~Mineev}(2012)}]{Mineev12}%
  \BibitemOpen
  \bibfield  {author} {\bibinfo {author} {\bibfnamefont {V.}~\bibnamefont {P.~Mineev}},\ }\bibfield  {title} {\bibinfo {title} {Whether there is the intrinsic hall effect in a multi-band superconductor?},\ }\href@noop {} {\bibfield  {journal} {\bibinfo  {journal} {Journal of the Physical Society of Japan}\ }\textbf {\bibinfo {volume} {81}},\ \bibinfo {pages} {093703} (\bibinfo {year} {2012})}\BibitemShut {NoStop}%
\bibitem [{\citenamefont {Wysoki\ifmmode~\acute{n}\else \'{n}\fi{}ski}\ \emph {et~al.}(2012)\citenamefont {Wysoki\ifmmode~\acute{n}\else \'{n}\fi{}ski}, \citenamefont {Annett},\ and\ \citenamefont {Gy\"orffy}}]{Gyroffy12}%
  \BibitemOpen
  \bibfield  {author} {\bibinfo {author} {\bibfnamefont {K.~I.}\ \bibnamefont {Wysoki\ifmmode~\acute{n}\else \'{n}\fi{}ski}}, \bibinfo {author} {\bibfnamefont {J.~F.}\ \bibnamefont {Annett}},\ and\ \bibinfo {author} {\bibfnamefont {B.~L.}\ \bibnamefont {Gy\"orffy}},\ }\bibfield  {title} {\bibinfo {title} {Intrinsic optical dichroism in the chiral superconducting state of ${\mathrm{sr}}_{2}{\mathrm{ruo}}_{4}$},\ }\href {https://doi.org/10.1103/PhysRevLett.108.077004} {\bibfield  {journal} {\bibinfo  {journal} {Phys. Rev. Lett.}\ }\textbf {\bibinfo {volume} {108}},\ \bibinfo {pages} {077004} (\bibinfo {year} {2012})}\BibitemShut {NoStop}%
\bibitem [{\citenamefont {Taylor}\ and\ \citenamefont {Kallin}(2013)}]{Kallin13}%
  \BibitemOpen
  \bibfield  {author} {\bibinfo {author} {\bibfnamefont {E.}~\bibnamefont {Taylor}}\ and\ \bibinfo {author} {\bibfnamefont {C.}~\bibnamefont {Kallin}},\ }\bibfield  {title} {\bibinfo {title} {Anomalous hall conductivity of clean sr2ruo4 at finite temperatures},\ }\href {https://doi.org/10.1088/1742-6596/449/1/012036} {\bibfield  {journal} {\bibinfo  {journal} {J. Phys.: Conf. Ser.}\ }\textbf {\bibinfo {volume} {449}},\ \bibinfo {pages} {012036} (\bibinfo {year} {2013})}\BibitemShut {NoStop}%
\bibitem [{\citenamefont {Denys}\ and\ \citenamefont {Brydon}(2021)}]{Brydon21}%
  \BibitemOpen
  \bibfield  {author} {\bibinfo {author} {\bibfnamefont {M.~D.~E.}\ \bibnamefont {Denys}}\ and\ \bibinfo {author} {\bibfnamefont {P.~M.~R.}\ \bibnamefont {Brydon}},\ }\bibfield  {title} {\bibinfo {title} {Origin of the anomalous hall effect in two-band chiral superconductors},\ }\href {https://doi.org/10.1103/PhysRevB.103.094503} {\bibfield  {journal} {\bibinfo  {journal} {Phys. Rev. B}\ }\textbf {\bibinfo {volume} {103}},\ \bibinfo {pages} {094503} (\bibinfo {year} {2021})}\BibitemShut {NoStop}%
\bibitem [{\citenamefont {Zhang}\ \emph {et~al.}(2024)\citenamefont {Zhang}, \citenamefont {Chen}, \citenamefont {Liu}, \citenamefont {Li}, \citenamefont {Wang},\ and\ \citenamefont {Huang}}]{Zhang24}%
  \BibitemOpen
  \bibfield  {author} {\bibinfo {author} {\bibfnamefont {J.-L.}\ \bibnamefont {Zhang}}, \bibinfo {author} {\bibfnamefont {W.}~\bibnamefont {Chen}}, \bibinfo {author} {\bibfnamefont {H.-T.}\ \bibnamefont {Liu}}, \bibinfo {author} {\bibfnamefont {Y.}~\bibnamefont {Li}}, \bibinfo {author} {\bibfnamefont {Z.}~\bibnamefont {Wang}},\ and\ \bibinfo {author} {\bibfnamefont {W.}~\bibnamefont {Huang}},\ }\bibfield  {title} {\bibinfo {title} {Quantum-geometry-induced anomalous hall effect in nonunitary superconductors and application to ${\mathrm{sr}}_{2}{\mathrm{ruo}}_{4}$},\ }\href {https://doi.org/10.1103/PhysRevLett.132.136001} {\bibfield  {journal} {\bibinfo  {journal} {Phys. Rev. Lett.}\ }\textbf {\bibinfo {volume} {132}},\ \bibinfo {pages} {136001} (\bibinfo {year} {2024})}\BibitemShut {NoStop}%
\bibitem [{\citenamefont {Hu}\ and\ \citenamefont {Huang}(2025)}]{Hu25}%
  \BibitemOpen
  \bibfield  {author} {\bibinfo {author} {\bibfnamefont {Y.-J.}\ \bibnamefont {Hu}}\ and\ \bibinfo {author} {\bibfnamefont {W.}~\bibnamefont {Huang}},\ }\bibfield  {title} {\bibinfo {title} {Quantum geometric superfluid weight in multiband superconductors: A microscopic interpretation},\ }\href {https://doi.org/10.1103/PhysRevB.111.134511} {\bibfield  {journal} {\bibinfo  {journal} {Phys. Rev. B}\ }\textbf {\bibinfo {volume} {111}},\ \bibinfo {pages} {134511} (\bibinfo {year} {2025})}\BibitemShut {NoStop}%
\bibitem [{\citenamefont {Bhattacharya}\ and\ \citenamefont {Black-Schaffer}(2026)}]{Bhattacharya26}%
  \BibitemOpen
  \bibfield  {author} {\bibinfo {author} {\bibfnamefont {A.}~\bibnamefont {Bhattacharya}}\ and\ \bibinfo {author} {\bibfnamefont {A.~M.}\ \bibnamefont {Black-Schaffer}},\ }\bibfield  {title} {\bibinfo {title} {Diamagnetic meissner response of odd-frequency superconducting pairing from quantum geometry},\ }\href {https://doi.org/10.1103/c57s-skv9} {\bibfield  {journal} {\bibinfo  {journal} {Phys. Rev. B}\ }\textbf {\bibinfo {volume} {113}},\ \bibinfo {pages} {094501} (\bibinfo {year} {2026})}\BibitemShut {NoStop}%
\bibitem [{\citenamefont {T\"orm\"a}(2023)}]{Torma23}%
  \BibitemOpen
  \bibfield  {author} {\bibinfo {author} {\bibfnamefont {P.}~\bibnamefont {T\"orm\"a}},\ }\bibfield  {title} {\bibinfo {title} {Essay: Where can quantum geometry lead us?},\ }\href {https://doi.org/10.1103/PhysRevLett.131.240001} {\bibfield  {journal} {\bibinfo  {journal} {Phys. Rev. Lett.}\ }\textbf {\bibinfo {volume} {131}},\ \bibinfo {pages} {240001} (\bibinfo {year} {2023})}\BibitemShut {NoStop}%
\bibitem [{\citenamefont {Zeng}\ \emph {et~al.}(2023)\citenamefont {Zeng}, \citenamefont {Xu}, \citenamefont {Wang},\ and\ \citenamefont {Hu}}]{Zeng23}%
  \BibitemOpen
  \bibfield  {author} {\bibinfo {author} {\bibfnamefont {M.}~\bibnamefont {Zeng}}, \bibinfo {author} {\bibfnamefont {D.-H.}\ \bibnamefont {Xu}}, \bibinfo {author} {\bibfnamefont {Z.-M.}\ \bibnamefont {Wang}},\ and\ \bibinfo {author} {\bibfnamefont {L.-H.}\ \bibnamefont {Hu}},\ }\bibfield  {title} {\bibinfo {title} {Spin-orbit coupled superconductivity with spin-singlet nonunitary pairing},\ }\href {https://doi.org/10.1103/PhysRevB.107.094507} {\bibfield  {journal} {\bibinfo  {journal} {Phys. Rev. B}\ }\textbf {\bibinfo {volume} {107}},\ \bibinfo {pages} {094507} (\bibinfo {year} {2023})}\BibitemShut {NoStop}%
\bibitem [{\citenamefont {Kane}\ and\ \citenamefont {Mele}(2005{\natexlab{a}})}]{KM05}%
  \BibitemOpen
  \bibfield  {author} {\bibinfo {author} {\bibfnamefont {C.~L.}\ \bibnamefont {Kane}}\ and\ \bibinfo {author} {\bibfnamefont {E.~J.}\ \bibnamefont {Mele}},\ }\bibfield  {title} {\bibinfo {title} {Quantum spin hall effect in graphene},\ }\href {https://doi.org/10.1103/PhysRevLett.95.226801} {\bibfield  {journal} {\bibinfo  {journal} {Phys. Rev. Lett.}\ }\textbf {\bibinfo {volume} {95}},\ \bibinfo {pages} {226801} (\bibinfo {year} {2005}{\natexlab{a}})}\BibitemShut {NoStop}%
\bibitem [{\citenamefont {Kane}\ and\ \citenamefont {Mele}(2005{\natexlab{b}})}]{KM052nd}%
  \BibitemOpen
  \bibfield  {author} {\bibinfo {author} {\bibfnamefont {C.~L.}\ \bibnamefont {Kane}}\ and\ \bibinfo {author} {\bibfnamefont {E.~J.}\ \bibnamefont {Mele}},\ }\bibfield  {title} {\bibinfo {title} {${Z}_{2}$ topological order and the quantum spin hall effect},\ }\href {https://doi.org/10.1103/PhysRevLett.95.146802} {\bibfield  {journal} {\bibinfo  {journal} {Phys. Rev. Lett.}\ }\textbf {\bibinfo {volume} {95}},\ \bibinfo {pages} {146802} (\bibinfo {year} {2005}{\natexlab{b}})}\BibitemShut {NoStop}%
\bibitem [{\citenamefont {Brydon}\ \emph {et~al.}(2019)\citenamefont {Brydon}, \citenamefont {Abergel}, \citenamefont {Agterberg},\ and\ \citenamefont {Yakovenko}}]{Brydon19}%
  \BibitemOpen
  \bibfield  {author} {\bibinfo {author} {\bibfnamefont {P.~M.~R.}\ \bibnamefont {Brydon}}, \bibinfo {author} {\bibfnamefont {D.~S.~L.}\ \bibnamefont {Abergel}}, \bibinfo {author} {\bibfnamefont {D.~F.}\ \bibnamefont {Agterberg}},\ and\ \bibinfo {author} {\bibfnamefont {V.~M.}\ \bibnamefont {Yakovenko}},\ }\bibfield  {title} {\bibinfo {title} {Loop currents and anomalous hall effect from time-reversal symmetry-breaking superconductivity on the honeycomb lattice},\ }\href {https://doi.org/10.1103/PhysRevX.9.031025} {\bibfield  {journal} {\bibinfo  {journal} {Phys. Rev. X}\ }\textbf {\bibinfo {volume} {9}},\ \bibinfo {pages} {031025} (\bibinfo {year} {2019})}\BibitemShut {NoStop}%
\bibitem [{\citenamefont {Faye}\ \emph {et~al.}(2015)\citenamefont {Faye}, \citenamefont {Sahebsara},\ and\ \citenamefont {S\'en\'echal}}]{Faye15}%
  \BibitemOpen
  \bibfield  {author} {\bibinfo {author} {\bibfnamefont {J.~P.~L.}\ \bibnamefont {Faye}}, \bibinfo {author} {\bibfnamefont {P.}~\bibnamefont {Sahebsara}},\ and\ \bibinfo {author} {\bibfnamefont {D.}~\bibnamefont {S\'en\'echal}},\ }\bibfield  {title} {\bibinfo {title} {Chiral triplet superconductivity on the graphene lattice},\ }\href {https://doi.org/10.1103/PhysRevB.92.085121} {\bibfield  {journal} {\bibinfo  {journal} {Phys. Rev. B}\ }\textbf {\bibinfo {volume} {92}},\ \bibinfo {pages} {085121} (\bibinfo {year} {2015})}\BibitemShut {NoStop}%
\bibitem [{\citenamefont {Ramires}(2022)}]{Ramires_2022}%
  \BibitemOpen
  \bibfield  {author} {\bibinfo {author} {\bibfnamefont {A.}~\bibnamefont {Ramires}},\ }\bibfield  {title} {\bibinfo {title} {Nonunitary superconductivity in complex quantum materials},\ }\href {https://doi.org/10.1088/1361-648X/ac6d3a} {\bibfield  {journal} {\bibinfo  {journal} {J. Phys.: Condens. Matter}\ }\textbf {\bibinfo {volume} {34}},\ \bibinfo {pages} {304001} (\bibinfo {year} {2022})}\BibitemShut {NoStop}%
\bibitem [{SM(2026)}]{SM}%
  \BibitemOpen
  \href@noop {} {\bibinfo {title} {See supplemental material for further details}} (\bibinfo {year} {2026})\BibitemShut {NoStop}%
\bibitem [{\citenamefont {Black-Schaffer}\ and\ \citenamefont {Doniach}(2007)}]{BS07}%
  \BibitemOpen
  \bibfield  {author} {\bibinfo {author} {\bibfnamefont {A.~M.}\ \bibnamefont {Black-Schaffer}}\ and\ \bibinfo {author} {\bibfnamefont {S.}~\bibnamefont {Doniach}},\ }\bibfield  {title} {\bibinfo {title} {Resonating valence bonds and mean-field $d$-wave superconductivity in graphite},\ }\href {https://doi.org/10.1103/PhysRevB.75.134512} {\bibfield  {journal} {\bibinfo  {journal} {Phys. Rev. B}\ }\textbf {\bibinfo {volume} {75}},\ \bibinfo {pages} {134512} (\bibinfo {year} {2007})}\BibitemShut {NoStop}%
\bibitem [{\citenamefont {Pathak}\ \emph {et~al.}(2010)\citenamefont {Pathak}, \citenamefont {Shenoy},\ and\ \citenamefont {Baskaran}}]{Bhaskaran10}%
  \BibitemOpen
  \bibfield  {author} {\bibinfo {author} {\bibfnamefont {S.}~\bibnamefont {Pathak}}, \bibinfo {author} {\bibfnamefont {V.~B.}\ \bibnamefont {Shenoy}},\ and\ \bibinfo {author} {\bibfnamefont {G.}~\bibnamefont {Baskaran}},\ }\bibfield  {title} {\bibinfo {title} {Possible high-temperature superconducting state with a $d+id$ pairing symmetry in doped graphene},\ }\href {https://doi.org/10.1103/PhysRevB.81.085431} {\bibfield  {journal} {\bibinfo  {journal} {Phys. Rev. B}\ }\textbf {\bibinfo {volume} {81}},\ \bibinfo {pages} {085431} (\bibinfo {year} {2010})}\BibitemShut {NoStop}%
\bibitem [{\citenamefont {Black-Schaffer}\ and\ \citenamefont {Honerkamp}(2014)}]{BS2014}%
  \BibitemOpen
  \bibfield  {author} {\bibinfo {author} {\bibfnamefont {A.~M.}\ \bibnamefont {Black-Schaffer}}\ and\ \bibinfo {author} {\bibfnamefont {C.}~\bibnamefont {Honerkamp}},\ }\bibfield  {title} {\bibinfo {title} {Chiral d-wave superconductivity in doped graphene},\ }\href@noop {} {\bibfield  {journal} {\bibinfo  {journal} {J. Phys.: Condens. Matter}\ }\textbf {\bibinfo {volume} {26}},\ \bibinfo {pages} {423201} (\bibinfo {year} {2014})}\BibitemShut {NoStop}%
\bibitem [{\citenamefont {Xu}\ \emph {et~al.}(2016)\citenamefont {Xu}, \citenamefont {Wessel},\ and\ \citenamefont {Meng}}]{Meng16}%
  \BibitemOpen
  \bibfield  {author} {\bibinfo {author} {\bibfnamefont {X.~Y.}\ \bibnamefont {Xu}}, \bibinfo {author} {\bibfnamefont {S.}~\bibnamefont {Wessel}},\ and\ \bibinfo {author} {\bibfnamefont {Z.~Y.}\ \bibnamefont {Meng}},\ }\bibfield  {title} {\bibinfo {title} {Competing pairing channels in the doped honeycomb lattice hubbard model},\ }\href {https://doi.org/10.1103/PhysRevB.94.115105} {\bibfield  {journal} {\bibinfo  {journal} {Phys. Rev. B}\ }\textbf {\bibinfo {volume} {94}},\ \bibinfo {pages} {115105} (\bibinfo {year} {2016})}\BibitemShut {NoStop}%
\bibitem [{\citenamefont {Huang}\ \emph {et~al.}(2024)\citenamefont {Huang}, \citenamefont {Zhou},\ and\ \citenamefont {Wang}}]{Wang24}%
  \BibitemOpen
  \bibfield  {author} {\bibinfo {author} {\bibfnamefont {J.}~\bibnamefont {Huang}}, \bibinfo {author} {\bibfnamefont {T.}~\bibnamefont {Zhou}},\ and\ \bibinfo {author} {\bibfnamefont {Z.~D.}\ \bibnamefont {Wang}},\ }\bibfield  {title} {\bibinfo {title} {Exotic topological phenomena in chiral superconducting states on doped quantum spin hall insulators with honeycomb lattices},\ }\href {https://doi.org/10.1103/PhysRevB.109.205144} {\bibfield  {journal} {\bibinfo  {journal} {Phys. Rev. B}\ }\textbf {\bibinfo {volume} {109}},\ \bibinfo {pages} {205144} (\bibinfo {year} {2024})}\BibitemShut {NoStop}%
\end{thebibliography}%

\newpage
\pagebreak

\setcounter{equation}{0}
\setcounter{figure}{0}

\onecolumngrid
\vspace{20pt}

\begin{center}
{\bf {\Large{Supplemental Material for \q{Quantum geometric anomalous Hall response in orbitally nonunitary superconductors}}}}
\end{center}

In this Supplemental Material (SM), we present the normal-state Hamiltonian in momentum space and also derive the anomalous Hall conductivity for both chiral $d$-wave spin-singlet and chiral $p$-wave equal-spin triplet pairing states, studied in the main text. 

\section{Normal-state Hamiltonian}
By Fourier transforming the real-space Hamiltonian introduced in the main text in Eq.~(1), the normal-state Hamiltonian in momentum space can be written as
\begin{align}
    H_N(\mk) = \sigma_0 \otimes \Big( \epsilon^x_\mk \tau_1 + \epsilon^y_\mk \tau_2 - \mu \tau_0 \Big) + t_2(\mk) \:\sigma_3 \otimes \tau_3\,, \label{eq:normal state Hamiltonian}
\end{align}
where $\sigma_i$ ($\tau_i$) are the Pauli matrices in spin (sublattice) space. The first two terms arises from nearest-neighbor hopping, the third term is the chemical potential and the last term is the symmetry allowed Kane-Mele spin-orbit coupling (SOC). The momentum dependent functions are given by
\begin{align}
    \epsilon^x_\mk = -t \sum_{j=1}^3 \cos (\mk\cdot\mathbf{R}_j), \hspace{10pt} \epsilon^y_\mk = t \sum_{j=1}^3 \sin (\mk\cdot\mathbf{R}_j),
\end{align}
\begin{figure}[h!]
    \centering
    \includegraphics[width=0.5\columnwidth]{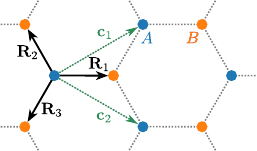}
    \caption{Schematic of the honeycomb lattice indicating the nearest-neighbor vectors $R_j$ and next-nearest-neighbor vectors $c_j$.}
    \label{fig:Fig1_SM}
\end{figure}
where $t$ is the nearest-neighbor hopping parameter and $\mathbf{R}_j$ are the nearest neighbor vectors connecting the two sites $A$ and $B$. The function describing the SOC term is given by
\begin{align}
    t_2(\mk) = -2\lambda_{\text{SOC}}[\sin(\mk\cdot\mathbf{c}_1) - \sin(\mk\cdot\mathbf{c}_2)],
\end{align}
where $\mathbf{c}_i$ are the vectors connecting next nearest neighbor sites. The nearest neighbor and next nearest neighbor bond vectors are illustrated in Fig.~\ref{fig:Fig1_SM}.
Since $t_2(\mk)$ is odd in $\mk$, the SOC term preserves both inversion and time-reversal symmetry, represented, respectively, by $\mathcal{I} = \sigma_0 \otimes \tau_1$, and $\mathcal{T} = i\sigma_2 \otimes \tau_0 \,\mathcal{K}$, where $\mathcal{K}$ stand for complex conjugation. It is easy to check that
\begin{align}
    \mathcal{I}H_N(\mk) \mathcal{I}^{-1} = H_N(-\mk), \hspace{10pt} \mathcal{T} H_N(\mk) \mathcal{T}^{-1} = H_N(-\mk),
\end{align}
and, consequently, the normal-state bands are spin degenerate. 

%START HERE
\section{Derivation of anomalous Hall conductivity}

\subsection{Opposite-spin pairing}\label{sec: opposite-spin pairing}
Within the Kubo linear response formalism, the anomalous Hall conductivity of a superconductor at optical frequency $\omega$ can be expressed, to one-loop order, similar to Eq.~(10-11) in the main text, as
\begin{align}
    \sigma_H(\omega)/e^2 = \frac{i}{4N \omega} \sum_{\mk}\frac{1}{\beta}\sum_{ i \nu_m} \mathrm{Tr}[\mathcal{V}^x_{\mk} \mathcal{G}_\mk(i\omega_n + i\nu_m) \mathcal{V}^y_{\mk}\mathcal{G}_\mk(i\nu_m) - (x \leftrightarrow y)]. \label{eq:hall conductivity startig point derivation spin-singlet}
\end{align}
where the limit $i\omega_n\to\omega+0^+$ is implied and must be performed after the Matsubara-frequency sum over $i \nu_m$. We proceed by first assuming a normal-state Hamiltonian that is block diagonal in spin, such as the model considered in the main text in Eq.~(1). For the superconducting pairing, we in this subsection consider pairing between opposite spins, primarily for spin-singlet pairing but the conclusion of the analysis and the derivations below also holds for opposite-spin triplet pairing. Then, the BdG Hamiltonian and the velocity matrix in the orbital basis can be expressed as
\begin{align}
    H_{\text{BdG}}(\mk) &= 
    \begin{pmatrix}
        H^\uparrow_N(\mk) & 0 & 0 &\Delta^{O}_\mk \\
        0 & H^\downarrow_N(\mk) & - \Delta^{O}_\mk &0 \\
        0 & -(\Delta^{O}_\mk)^\dagger & -H_N^{\uparrow *}(\bar{\mk}) & 0 \\
        (\Delta^{O}_\mk)^\dagger & 0 & 0 & -H_N^{\downarrow *}(\bar{\mk})
    \end{pmatrix},\label{eq: SM_HBdG1} \\
    (\mathcal{V}^{\mu}_{\mk})^\text{O} &= 
    \begin{pmatrix}
        V^{O\uparrow}_{\mu\mk}& & & \\
        & V^{O\downarrow}_{\mu\mk} & & \\
        & & -V^{O\uparrow T}_{\mu\Bar{\mk}} & \\
        & & &-V^{O\downarrow T}_{\mu\Bar{\mk}}
        \end{pmatrix},
\end{align}
where we use the superscript $O$ to denote that these matrices are written in the orbital (sublattice) basis, and $\bar{\mk} = -\mk$. We further use that $(\Delta^O_\mk)^{\uparrow\downarrow} = -(\Delta^O_\mk)^{\downarrow \uparrow} \equiv \Delta^O_\mk$ for spin-singlet pairing. For opposite-spin triplet pairing, the pairing matrix is instead even when changing the spin indices.  
Note that the entries in Eq.~(\ref{eq: SM_HBdG1}) are $2\times 2$ blocks in orbital space. Since the normal-state Hamiltonian is block diagonal in spin, we transform the BdG Hamiltonian and the velocity matrix to the band basis by the following unitary transformation:
\begin{align}
    \mathcal{W}_\mk^\dagger = 
    \begin{pmatrix}
        \mathcal{U}^\dagger_{\mk\uparrow} & & & \\
        & \mathcal{U}^\dagger_{\mk\downarrow} & & \\
        & & \mathcal{U}^T_{\bar{\mk}\uparrow} & \\
        & & & \mathcal{U}^T_{\bar{\mk}\downarrow}
    \end{pmatrix},
\end{align}
such that in the band basis, BdG Hamiltonian $\Tilde{H}_{BdG}(\mk) = \mathcal{W}^\dagger_\mk H_{BdG}(\mk) \mathcal{W}_\mk$ and the velocity matrix $\tilde{\mathcal{V}}^\mu_{\mk} = \mathcal{W}^\dagger_\mk (\mathcal{V}^{\mu}_{\mk})^\text{O} \mathcal{W}_\mk$ are
\begin{align}
    \Tilde{H}_{BdG}(\mk) = 
    \begin{pmatrix}
        \Gamma_{\mk\uparrow} & 0 & 0 & \Delta_\mk \\
        0 & \Gamma_{\mk\downarrow}  & -\Delta_\mk & 0 \\
        0 & -\Delta^\dagger_\mk &  -\Gamma_{\bar{\mk}\uparrow} & 0 \\
        \Delta^\dagger_\mk & 0 & 0 & -\Gamma_{\bar{\mk}\downarrow}
    \end{pmatrix}, \hspace{5pt}
    \tilde{\mathcal{V}}^\mu_{\mk} = 
    \begin{pmatrix}
        V_{\mu\mk\uparrow}& & & \\
        & V_{\mu\mk\downarrow} & & \\
        & & -V^{ T}_{\mu\Bar{\mk}\uparrow} & \\
        & & &-V^{T}_{\mu\Bar{\mk}\downarrow}
    \end{pmatrix}.
\end{align}
Here, $\Gamma_{\mk s}  = \text{diag}(\epsilon_{\mk 1 s}, \epsilon_{\mk 2 s})$ with $\epsilon_{\mk i s}$ denotes the normal-state band and $\Delta_\mk = \mathcal{U}^\dagger_{\mk \uparrow} \Delta^{O}(\mk) \mathcal{U}^*_{\bar{\mk} \downarrow}$ is the pairing matrix in the band basis. The BdG Hamiltonian can further be made block diagonal by a unitary transformation, resulting in:
\begin{align}
    \tilde{H}^\prime_{BdG}(\mk) =  
    \begin{pmatrix}
        \Tilde{H}^{\uparrow\downarrow}_{BdG}(\mk) & 0 \\
        0 & \Tilde{H}^{\downarrow\uparrow}_{BdG}(\mk)
    \end{pmatrix},
    \hspace{10pt}
    \tilde{\mathcal{V}}^{\mu \prime}_{\mk} = 
    \begin{pmatrix}
        \mathcal{V}^{\uparrow\downarrow}_{\mu\mk} & 0 \\
        0 & \mathcal{V}^{\downarrow\uparrow}_{\mu\mk}
    \end{pmatrix}
\end{align}
with
\begin{align}
    \Tilde{H}_{BdG}^{\uparrow\downarrow}(\mk) = 
    \begin{pmatrix}
        \Gamma_{\mk \uparrow} & \Delta_\mk \\
        \Delta^\dagger_\mk & -\Gamma_{\bar{\mk} \downarrow}
    \end{pmatrix},
    \hspace{10 pt}
    \mathcal{V}^{\uparrow\downarrow}_{\mu\mk} = 
    \begin{pmatrix}
        V_{\mu\mk \uparrow} &  \\
         & -V^T_{\mu\Bar{\mk}\downarrow}
    \end{pmatrix}.\label{eq:bdg spin singlet blocks}
\end{align}
The other blocks are related by the exchange of spin: $\uparrow \leftrightarrow \downarrow$, with the understanding that $\Delta_\mk = \Delta_\mk^{\uparrow\downarrow} \to \Delta^{\downarrow\uparrow}_\mk = - \Delta^{\uparrow\downarrow}_\mk = -\Delta_\mk$. In this basis, the Green's function is also block diagonal $\mathcal{G}_{\mk,i\nu_m} = (i\nu_m - \tilde{H}^\prime_{BdG})^{-1} = \text{diag}(\mathcal{G}^{\uparrow\downarrow}, \mathcal{G}^{\downarrow\uparrow})$. The Hall conductivity in Eq.~\eqref{eq:hall conductivity startig point derivation spin-singlet} then simplifies to $\sigma_H(\omega) = \sigma_H^{\uparrow\downarrow}(\omega) + \sigma_H^{\downarrow\uparrow}(\omega)$, 
with
\begin{align}
    \sigma^{\uparrow\downarrow}_H(\omega) = \frac{ie^2}{4N\beta\omega} \sum_{\mk\, i\nu_m} \mathrm{Tr}[\mathcal{V}^{\uparrow\downarrow}_{x\mk} \mathcal{G}^{\uparrow\downarrow}_\mk(i\omega_n + i\nu_m) \mathcal{V}^{\uparrow\downarrow}_{y\mk}\mathcal{G}^{\uparrow\downarrow}_\mk(i\nu_m) - (x \leftrightarrow y)], \label{eq: sigma_updown general}
\end{align}
and $\sigma^{\downarrow\uparrow}(\omega)$ being related to $\sigma^{\uparrow\downarrow}_H(\omega)$ by exchanging  $\uparrow\leftrightarrow\downarrow$. If we express Eq.~\eqref{eq: sigma_updown general} in the BdG eigenbasis, and perform the Matsubara summation, we obtain
\begin{align}
    \sigma^{\uparrow\downarrow}_H(\omega) = \frac{ie^2}{4N\omega} \sum_{\mk, ab}\frac{n_F(E_{a\mk}) - n_F(E_{b\mk})}{\omega + i0^+ + E_{a\mk} - E_{b\mk}} \Big[ \bra{b\mk} \mathcal{V}^{\uparrow\downarrow}_{x\mk} \ket{a\mk}\bra{a\mk} \mathcal{V}^{\uparrow\downarrow}_{y\mk}  \ket{b\mk} - (x \leftrightarrow y) \Big], \label{eq:hall conductivity block spin singlet}
\end{align}
where $n_F(E)$ is the Fermi-Dirac distribution, $E_{a\mk}$ are the eigenvalues and $\ket{a\mk}$ the eigenvectors of the BdG Hamiltonian $\tilde{H}_{\text{BdG}}^{\uparrow \downarrow}$.

We note that the pairing matrix generally contains both intraband and interband pairing terms after the unitary transformation from the orbital basis to the band basis. Since we are interested in the intraband contributions to the Hall conductivity, we explicitly put the interband pairing to zero by hand, which gives us the Hamiltonian:
\begin{align}
    \Tilde{H}^{\uparrow\downarrow}_{BdG}(\mk) = 
    \begin{pmatrix}
        \epsilon_{\mk 1\uparrow} & 0 & \Delta_{\mk 1} & 0 \\
        0 & \epsilon_{\mk 2\uparrow} & 0 & \Delta_{\mk 2} \\
        \Delta^*_{\mk 1} & 0 & -\epsilon_{\bar{\mk} 1\downarrow} & 0 \\
        0 & \Delta^*_{\mk 2} & 0 & -\epsilon_{\bar{\mk} 2\downarrow}
    \end{pmatrix}= \begin{pmatrix}
        \epsilon_{\mk 1\uparrow} & 0 & \Delta_{\mk 1} & 0 \\
        0 & \epsilon_{\mk 2\uparrow} & 0 & \Delta_{\mk 2} \\
        \Delta^*_{\mk 1} & 0 & -\epsilon_{\mk 1\uparrow} & 0 \\
        0 & \Delta^*_{\mk 2} & 0 & -\epsilon_{\mk 2\uparrow}
    \end{pmatrix},
\end{align}
where we have used $\Gamma_{\bar{\mk}\downarrow} = \Gamma_{\mk\uparrow}$ due to TRS in the normal state. Since this Hamiltonian effectively describes two separate conventional BCS superconductors, the eigenvectors takes a simple form
\begin{comment}
\begin{align}
    \ket{1\mk\uparrow} &= (u_{\mk1\uparrow}, 0, v_{\mk1\uparrow}, 0)^T, 
    \hspace{10pt} &&\Tilde{H}^{\uparrow\downarrow}_{BdG} \ket{1\mk\uparrow} = E_{\mk 1 \uparrow} \ket{1\mk\uparrow}, \label{eq:zhang section bdg eigenvector 1 first} \\
    \ket{\Bar{1}\mk\uparrow} &= (-v^*_{\mk1\uparrow}, 0, u^*_{\mk1\uparrow}, 0)^T, 
    \hspace{10pt} &&\Tilde{H}^{\uparrow\downarrow}_{BdG} \ket{\Bar{1}\mk\uparrow} = -E_{\mk 1 \uparrow} \ket{\Bar{1}\mk\uparrow}, \\
    \ket{2\mk\uparrow} &= (0, u_{\mk2\uparrow}, 0, v_{\mk2\uparrow})^T,
    \hspace{10pt} &&\Tilde{H}^{\uparrow\downarrow}_{BdG} \ket{2\mk\uparrow} = E_{\mk 2 \uparrow} \ket{2\mk\uparrow}, \\
    \ket{\Bar{2}\mk\uparrow} &= (0, -v^*_{\mk2\uparrow}, 0, u^*_{\mk2\uparrow})^T, 
    \hspace{10pt} &&\Tilde{H}^{\uparrow\downarrow}_{BdG} \ket{\Bar{2}\mk\uparrow} = -E_{\mk 2 \uparrow} \ket{\Bar{2}\mk\uparrow}, \label{eq:zhang section bdg eigenvector 1 last}
\end{align}
\end{comment}
\begin{align}
    \ket{1\mk} &= (u_{\mk1}, 0, v_{\mk}, 0)^T, 
    \hspace{10pt} &&\Tilde{H}^{\uparrow\downarrow}_{BdG} \ket{1\mk} = E_{\mk 1} \ket{1\mk}, \label{eq:zhang section bdg eigenvector 1 first} \\
    \ket{\Bar{1}\mk} &= (-v^*_{\mk1 }, 0, u^*_{\mk1}, 0)^T, 
    \hspace{10pt} &&\Tilde{H}^{\uparrow\downarrow}_{BdG} \ket{\Bar{1}\mk } = -E_{\mk 1 } \ket{\Bar{1}\mk}, \\
    \ket{2\mk} &= (0, u_{\mk2}, 0, v_{\mk2})^T,
    \hspace{10pt} &&\Tilde{H}^{\uparrow\downarrow}_{BdG} \ket{2\mk} = E_{\mk 2 } \ket{2\mk}, \\
    \ket{\Bar{2}\mk} &= (0, -v^*_{\mk2 }, 0, u^*_{\mk2 })^T, 
    \hspace{10pt} &&\Tilde{H}^{\uparrow\downarrow}_{BdG} \ket{\Bar{2}\mk } = -E_{\mk 2 } \ket{\Bar{2}\mk}, \label{eq:zhang section bdg eigenvector 1 last}
\end{align}
where
\begin{align}
    \abs{u_{\mk i }}^2 = \frac{1}{2}\left( 1 + \frac{\epsilon_{\mk i }}{E_{\mk i }}\right), \hspace{10pt} \abs{v_{\mk i }}^2 = \frac{1}{2}\left( 1 - \frac{\epsilon_{\mk i }}{E_{\mk i }}\right), \quad \text{and} \quad E_{\mk i} = \sqrt{\epsilon_{\mk i }^2 + \abs{\Delta_{\mk i}}^2}. \label{eq:eigenvalues and coherence factors opposite pairing}
\end{align}
We now evaluate Eq.~\eqref{eq:hall conductivity block spin singlet} at zero temperature, where the Fermi function becomes a step function and the only allowed processes (virtual excitations) starts from an occupied state below the Fermi energy $\ket{\Bar{i} \mk}$, going to a unoccupied state above the Fermi energy $\ket{j \mk }$ and back.  We can then write Eq.~\eqref{eq:hall conductivity block spin singlet} as
\begin{align}
     \sigma^{\uparrow\downarrow}_H&(\omega) =\frac{ie^2}{4N\omega} \sum_{\mk} \nonumber \\
     &\times \Bigg( \frac{1}{\omega + i0^+ + E_{\mk \Bar{1}} - E_{\mk 2}} \Big[ \bra{2\mk} \mathcal{V}^{\uparrow\downarrow}_{x\mk} \ket{\Bar{1}\mk}\bra{\Bar{1}\mk} \mathcal{V}^{\uparrow\downarrow}_{y\mk}  \ket{2\mk} - (x \leftrightarrow y) \Big] - (\Bar{1} \leftrightarrow 2) \Bigg) + (1 \leftrightarrow 2), \label{eq:sigma_H updown half way there}
\end{align}
where $(\Bar{1} \leftrightarrow 2)$ means exchanging $\bar{1}$ with $2$, which accounts for the second Fermi-Dirac distribution in Eq.~(\ref{eq:hall conductivity block spin singlet}). This is in contrast to $(1 \leftrightarrow 2)$, which accounts for processes starting from $\bar{2}$ to $1$ and back, where all 1's and 2's are exchanged, i.e., $\bar{1} \rightarrow \bar{2}$, and $2 \rightarrow 1$.
We note here that transitions within the same band $i \leftrightarrow \bar{i}$ do not contribute. We see this by explicitly evaluating
\begin{align}
    \bra{i\mk} \mathcal{V}^{\uparrow\downarrow}_{x\mk} \ket{\Bar{i}\mk}\bra{\Bar{i}\mk} \mathcal{V}^{\uparrow\downarrow}_{y\mk}  \ket{i\mk} - (x \leftrightarrow y) = |u_{\mk i }|^2 |v_{\mk i }|^2(V_{x\mk \uparrow}^{ii} + (V_{x\bar{\mk}\downarrow}^{T})^{ii})(V_{y\mk \uparrow}^{ii} + (V_{y\bar{\mk}\downarrow}^{T})^{ii} ) - (x \leftrightarrow y) = 0. \nonumber
\end{align}
By using $E_{ \mk \bar1} = - E_{\mk 1 }$, Eq.~\eqref{eq:sigma_H updown half way there} can be simplified to
\begin{align}
     &\sigma^{\uparrow\downarrow}_H(\omega) = \frac{ie^2}{2N} \sum_{\mk}\Bigg( \frac{1}{(\omega + i0^+)^2 - (E_{\mk \Bar{1}} - E_{\mk 2})^2} \Big[ \bra{2\mk} \mathcal{V}^{\uparrow\downarrow}_{x\mk} \ket{\Bar{1}\mk}\bra{\Bar{1}\mk} \mathcal{V}^{\uparrow\downarrow}_{y\mk}  \ket{2\mk} - (x \leftrightarrow y) \Big] \Bigg) + (1 \leftrightarrow 2) \nonumber \\
     &= \frac{ie^2}{2N} \sum_{\mk} \frac{1}{(\omega + i0^+)^2 - (E_{\mk 1} + E_{\mk 2})^2} \Bigg( \Big[ \bra{2\mk} \mathcal{V}^{\uparrow\downarrow}_{x\mk} \ket{\Bar{1}\mk}\bra{\Bar{1}\mk} \mathcal{V}^{\uparrow\downarrow}_{y\mk}  \ket{2\mk} - (x \leftrightarrow y) \Big] + (1 \leftrightarrow 2) \Bigg).  \label{eq:call cond almost final singlet}
\end{align}
The matrix elements are evaluated as
\begin{align}
    &\bra{2\mk} \mathcal{V}^{\uparrow\downarrow}_{x\mk}  \ket{\Bar{1}\mk}\bra{\Bar{1}\mk} \mathcal{V}^{\uparrow\downarrow}_{y\mk} \ket{2\mk} - (x \leftrightarrow y) = \nonumber \\
    &= \abs{u_{\mk2}}^2\abs{v_{\mk1}}^2 (V_{x\mk \uparrow}^{21}V_{y\mk \uparrow}^{12} - V_{x\mk \uparrow}^{12}V_{y\mk \uparrow}^{21})
    + \abs{u_{\mk1}}^2\abs{v_{\mk2}}^2 ((V_{x\bar{\mk} \downarrow}^T)^{21}(V_{y\bar{\mk} \downarrow}^T)^{12} -(V_{x\bar{\mk} \downarrow}^T)^{12}(V_{y\bar{\mk} \downarrow}^T)^{21} ) \nonumber \\
    &+ v_{\mk1}u_{\mk2} u_{\mk1}^*v_{\mk2}^*((V_{x\bar{\mk} \downarrow}^T)^{21}V_{y\mk \uparrow}^{12} - V_{x\mk \uparrow}^{12}(V_{y\bar{\mk} \downarrow}^T)^{21}) + u_{\mk 1}v_{\mk 2}v_{\mk 1}^*u_{\mk 2}^*(V_{x\mk \uparrow}^{21}(V_{y\bar{\mk} \downarrow}^T)^{12}- (V_{x\bar{\mk} \downarrow}^T)^{12}V_{y\mk \uparrow}^{21}).\nonumber
\end{align}
and adding $(1 \leftrightarrow 2)$ results in
\begin{align}
    \Big[ \bra{2\mk} \mathcal{V}^{\uparrow\downarrow}_{x\mk}  &\ket{\Bar{1}\mk} \bra{\Bar{1}\mk} \mathcal{V}^{\uparrow\downarrow}_{y\mk} \ket{2\mk} - (x \leftrightarrow y) \Big] + (1 \leftrightarrow 2) \nonumber \\
    &=  (\abs{u_{\mk 1}}^2\abs{v_{\mk 2}}^2 - \abs{u_{\mk 2}}^2\abs{v_{\mk 1}}^2)(Q^{e,\uparrow}_{xy} -Q^{h,\downarrow}_{xy} ) = \frac{1}{2}\left( \frac{\epsilon_{\mk 1}}{E_{\mk 1}} - \frac{\epsilon_{\mk 2}}{E_{\mk 2}} \right) (Q^{e,\uparrow}_{xy} -Q^{h,\downarrow}_{xy} ), \label{eq:inner product coeff}
\end{align}
where we have introduced
\begin{align}
    Q^{e, s}_{xy} = V_{x\mk s}^{12}V_{y\mk s}^{21} - V_{y\mk s}^{12}V_{x\mk s}^{21}, \hspace{10pt} Q^{h, s}_{xy} = (V_{x\bar{\mk}s}^T)^{12}(V_{y\bar{\mk}s}^T)^{21} - (V_{y\bar{\mk}s}^T)^{12}(V_{x\bar{\mk}s}^T)^{21}, \label{eq:Qe and Qh def}
\end{align}
as in the main text. Inserting Eq.~\eqref{eq:inner product coeff} in Eq.~\eqref{eq:call cond almost final singlet}, the Hall conductivity becomes
\begin{align}
     \sigma^{\uparrow\downarrow}_H(\omega) 
     = \frac{ie^2}{4N} \sum_{\mk} \frac{\epsilon_{\mk 1}/E_{\mk 1} - \epsilon_{\mk 2}/E_{\mk 2}}{(\omega + i0^+)^2 - (E_{\mk 1} + E_{\mk 2})^2}(Q^{e, \uparrow}_{xy} -Q^{h, \downarrow}_{xy} ). \label{eq:haaaall singlet}
\end{align}

We next note that $Q^{e, s}_{xy}$ and $Q^{h, s}_{xy}$ are defined in terms of the normal state velocity matrix in the band basis. They can therefore be written in terms of the spin Berry curvature~\cite{Zhang24}
\begin{align}
    \Omega^s_{1,xy}(\mk) = i \frac{V_{x\mk s}^{12}V_{y\mk s}^{21} - V_{y\mk s}^{12}V_{x\mk s}^{21}}{(\epsilon_{1\mk s} - \epsilon_{2\mk s})^2}\,,
\end{align}
as
\begin{align}
    Q^{e, \uparrow}_{xy} &= V_{x\mk\uparrow}^{12}V_{y\mk\uparrow}^{21} - V_{y\mk\uparrow}^{12}V_{x\mk\uparrow}^{21} = -i (\epsilon_{1\mk\uparrow} -  \epsilon_{2\mk\uparrow})^2 \Omega^\uparrow_{1,xy}(\mk)\,, \\
    Q^{h, \downarrow}_{xy} &= (V_{x\bar{\mk}\downarrow}^T)^{12}(V_{y\bar{\mk}\downarrow}^T)^{21} - (V_{y\bar{\mk}\downarrow}^T)^{12}(V_{x\bar{\mk}\downarrow}^T)^{21} =  - \Big[V_{x\bar{\mk}\downarrow}^{12}V_{y\bar{\mk}\downarrow}^{21} - V_{y\bar{\mk}\downarrow}^{12}V_{x\bar{\mk}\downarrow}^{21} \Big] \\
    &= i(\epsilon_{1\bar{\mk}\downarrow} - \epsilon_{2\bar{\mk}\downarrow})^2\Omega^\downarrow_{1,xy}(\bar{\mk}) = -i (\epsilon_{1\mk\uparrow} -  \epsilon_{2\mk\uparrow})^2 \Omega^\uparrow_{1,xy}(\mk)\,.
\end{align}
Here we again use $\epsilon_{i\bar{\mk}\downarrow}=\epsilon_{i\mk\uparrow}$ and $\Omega^\downarrow_{1,xy}(\bar{\mk}) = - \Omega^\uparrow_{1,xy}(\mk)$, due to the assumed TRS in the normal state $H_N(\mathbf{k})$ in Eq.(\ref{eq: SM_HBdG1}) and we also use $\Omega^\uparrow_{1,xy}(\mk) = - \Omega^\uparrow_{2,xy}(\mk)$. We then see that $Q^{e, \uparrow}_{xy} = Q^{h, \downarrow}_{xy}$, and it follows that the Hall conductivity in Eq.~\eqref{eq:haaaall singlet} is zero. Note that $\sigma_{\downarrow\uparrow}$ vanished for the same reasons, by $\uparrow\leftrightarrow\downarrow$. We also note that this conclusion remains valid regardless of whether the BdG eigenstates are spin split (due to interband pairing). 

We thus reach the conclusion that for a superconducting state with a spin-singlet order parameter, without interband pairing, the Hall conductivity remains zero if TRS is present in the normal state:
\begin{align}
    \sigma_H(\omega) = \sigma_{\uparrow\downarrow}(\omega) + \sigma_{\downarrow\uparrow}(\omega) = 0\,,
\end{align}
This proves the statements in the main text about zero AHR for spin-singlet intraband pairing.
The same conclusion also holds for the $m_z=0$ spin-triplet pairing state. 

\subsection{Equal-spin triplet pairing}\label{sec: equal spin-triplet pairing}
It is straight forward to generalize the above derivation for spin-triplet pairing between equal spin
\begin{align}
    \Delta^O_\mk = 
    \begin{pmatrix}
        (\Delta_\mk^{O})^{\uparrow} & 0 \\
        0 & (\Delta^{O}_\mk)^\downarrow
    \end{pmatrix},
\end{align}
The BdG Hamiltonian and velocity matrix in Nambu space can again be written in a block diagonal form in the band basis
\begin{align}
    \tilde{H}^\prime_{BdG}(\mk) =  
    \begin{pmatrix}
        \Tilde{H}^\uparrow_{BdG}(\mk) & 0 \\
        0 & \Tilde{H}^\downarrow_{BdG}(\mk)
    \end{pmatrix},
    \hspace{10pt}
    \tilde{\mathcal{V}}^\mu_{\mk} = 
    \begin{pmatrix}
        \mathcal{V}^\uparrow_{\mu\mk} & 0 \\
        0 & \mathcal{V}^\downarrow_{\mu\mk}
    \end{pmatrix},
\end{align}
where now
\begin{align}
    \Tilde{H}_{BdG}^s(\mk) = 
    \begin{pmatrix}
        \Gamma_{\mk s} & \Delta_{\mk s} \\
        \Delta^\dagger_{\mk s} & -\Gamma_{\bar{\mk} s}
    \end{pmatrix},
    \hspace{10 pt}
    \mathcal{V}^s_{\mu\mk} = 
    \begin{pmatrix}
        V_{\mu\mk s} &  \\
         & -V^T_{\mu\Bar{\mk}s}
    \end{pmatrix}.
\end{align}
The Hall conductivity becomes $\sigma_H(\omega) = \sigma_H^{\uparrow \uparrow}(\omega) + \sigma_H^{\downarrow \downarrow}(\omega)$ with
\begin{align}
    \sigma^{\uparrow\uparrow}_H(\omega) = \frac{ie^2}{4N\beta\omega} \sum_{\mk\, i \nu_m} \Tr[\mathcal{V}^\uparrow_{x\mk} \mathcal{G}^\uparrow_\mk(i\omega_n + i \nu_m) \mathcal{V}^\uparrow_{y\mk}\mathcal{G}^\uparrow_\mk(i \nu_m) - (x \leftrightarrow y)],
\end{align}
and $\sigma^{\downarrow\downarrow}$ related by $\uparrow\leftrightarrow\downarrow$. After performing the Matsubara summation, the expression becomes
\begin{align}
    \sigma^{\uparrow\uparrow}_H(\omega)/e^2 = \frac{i}{4N\omega} \sum_{\mk ab}\frac{n_F(E_{a\mk\uparrow}) - n_F(E_{b\mk\uparrow})}{\omega + i0^+ + E_{a\mk\uparrow} - E_{b\mk\uparrow}} \Big[ \bra{b\mk \uparrow} \mathcal{V}^\uparrow_{x\mk} \ket{a\mk\uparrow}\bra{a\mk \uparrow} \mathcal{V}^\uparrow_{y\mk}  \ket{b\mk\uparrow} - (x \leftrightarrow y) \Big], \label{eq:spin up hall conductivity}
\end{align}
where $E_{a\mk\uparrow}$ and $\ket{a\mk\uparrow}$ are now the eigenvalues and eigenvectors of the spin-resolved BdG Hamiltonian
\begin{align}
    \Tilde{H}^\uparrow_{BdG}(\mk) = 
    \begin{pmatrix}
        \epsilon_{\mk 1\uparrow} & 0 & \Delta_{\mk 1 \uparrow} & 0 \\
        0 & \epsilon_{\mk 2\uparrow} & 0 & \Delta_{\mk 2 \uparrow} \\
        \Delta^*_{\mk 1 \uparrow} & 0 & -\epsilon_{\bar{\mk} 1\uparrow} & 0 \\
        0 & \Delta^*_{\mk 2 \uparrow} & 0 & -\epsilon_{\bar{\mk} 2\uparrow}
    \end{pmatrix}, \label{eq:spin respolved BdG Hamiltonian equal-spin triplet}
\end{align}
where we again have by hand enforced vanishing interband pairing. 
We can write down the eigenenergies and eigenvectors of the Hamiltonian in Eq.~\eqref{eq:spin respolved BdG Hamiltonian equal-spin triplet} (for simplicity we keep the same notation as in Section \ref{sec: opposite-spin pairing}, although these energies and vectors are different)
\begin{align}
    \ket{1\mk\uparrow} &= (u_{\mk1\uparrow}, 0, v_{\mk1\uparrow}, 0)^T, 
    \hspace{10pt} &&\Tilde{H}^\uparrow_{BdG} \ket{1\mk\uparrow} = E_{\mk 1 \uparrow} \ket{1\mk\uparrow}, \label{eq:zhang section bdg eigenvector first} \\
    \ket{\Bar{1}\mk\uparrow} &= (-v^*_{\mk1\uparrow}, 0, u^*_{\mk1\uparrow}, 0)^T, 
    \hspace{10pt} &&\Tilde{H}^\uparrow_{BdG} \ket{\Bar{1}\mk\uparrow} = E_{\mk \bar{1} \uparrow} \ket{\Bar{1}\mk\uparrow}, \\
    \ket{2\mk\uparrow} &= (0, u_{\mk2\uparrow}, 0, v_{\mk2\uparrow})^T,
    \hspace{10pt} &&\Tilde{H}^\uparrow_{BdG} \ket{2\mk\uparrow} = E_{\mk 2 \uparrow} \ket{2\mk\uparrow}, \\
    \ket{\Bar{2}\mk\uparrow} &= (0, -v^*_{\mk2\uparrow}, 0, u^*_{\mk2\uparrow})^T, 
    \hspace{10pt} &&\Tilde{H}^\uparrow_{BdG} \ket{\Bar{2}\mk\uparrow} = E_{\mk \bar{2} \uparrow} \ket{\Bar{2}\mk\uparrow}, \label{eq:zhang section bdg eigenvector last}
\end{align}
with
\begin{align}
    E_{\mk i \uparrow} = \delta_{i\mk\uparrow} + \sqrt{\xi_{i\mk \uparrow}^2 + |\Delta_{i\mk \uparrow}|^2}, \hspace{10pt} E_{\mk \bar{i} \uparrow} = \delta_{i\mk\uparrow} - \sqrt{\xi_{i\mk \uparrow}^2 + |\Delta_{i\mk \uparrow}|^2}, \label{eq:BdG eigenvalues no IS}
\end{align}
where we define $\delta_{i\mk\uparrow} = (\epsilon_{i\mk\uparrow}- \epsilon_{i\bar{\mk}\uparrow})/2$ and $\xi_{i\mk\uparrow} = (\epsilon_{i\mk\uparrow} + \epsilon_{i\bar{\mk}\uparrow})/2$. We choose the gauge such that
\begin{align}
    u_{\mk i\uparrow} = \sqrt{\frac{1}{2}\Biggl( 1 + \frac{\xi_{\mk i\uparrow}}{\sqrt{\xi_{i\mk \uparrow}^2 + |\Delta_{i\mk \uparrow}|^2}}\Biggr)}, \hspace{10pt} v_{\mk i\uparrow} = e^{-i\phi_{\mk i}}\sqrt{\frac{1}{2}\Biggl( 1 - \frac{\xi_{\mk i\uparrow}}{\sqrt{\xi_{i\mk \uparrow}^2 + |\Delta_{i\mk \uparrow}|^2}}\Biggr)},
\end{align}
with $e^{-i\phi_{\mk i}} = \Delta_{\mk i}^*/|\Delta_{\mk i}|$. 
 
If we now assume a normal state with inversion symmetry, we have $\epsilon_{\mk i\uparrow} = \epsilon_{\bar{\mk} i\uparrow} \equiv \epsilon_{i\mk}$, $\delta_{i\mk\uparrow} = 0$ and $\xi_{\mk i\uparrow} = \epsilon_{i\mk}$. We can then follow a similar procedure as in Sec.~\ref{sec: opposite-spin pairing}, resulting in the spin-resolved Hall conductivity
\begin{align}
\label{eq:sigmaHupup}
     \sigma^{\uparrow\uparrow}_H(\omega) 
     = \frac{ie^2}{4N} \sum_{\mk} \frac{\epsilon_{1\mk}/E_{1\mk\uparrow} - \epsilon_{2\mk}/E_{2\mk\uparrow}}{(\omega + i0^+)^2 - (E_{1\mk\uparrow} + E_{2\mk\uparrow})^2}(Q^{e,\uparrow}_{xy} - Q^{h,\uparrow}_{xy}),
\end{align}
which is the expression used in Eq.~(20) in the main text where $E_{a\mk s} = \sqrt{\epsilon_{a\mk} + |\Delta_{\mk s}|^2}$ are the BdG eigenvalues and
\begin{align}
    Q^{h,\uparrow}_{xy} =&(V_{x\bar{\mk}\uparrow}^T)^{12}(V_{y\bar{\mk}\uparrow}^T)^{21} - (V_{y\bar{\mk}\uparrow}^T)^{12}(V_{x\bar{\mk}\uparrow}^T)^{21} =  - \Big[V_{x\bar{\mk}\uparrow}^{12}V_{y\bar{\mk}\uparrow}^{21} - V_{y\bar{\mk}\uparrow}^{12}V_{x\bar{\mk}\uparrow}^{21} \Big] \nonumber \\
    &= i(\epsilon_{1\bar{\mk}\uparrow} - \epsilon_{2\bar{\mk}\uparrow})^2\Omega^\uparrow_{1,xy}(\bar{\mk}) = i (\epsilon_{1\mk} -  \epsilon_{2\mk})^2 \Omega^\uparrow_{1,xy}(\mk)\,.
\end{align}
In the last equality we used that $\Omega^\uparrow_{1,xy}(\bar{\mk}) =  \Omega^\uparrow_{1,xy}(\mk)$ if inversion symmetry is present in the normal state. In contrast to the opposite-spin pairing case, here $Q^{h,\uparrow}_{xy} = - Q^{e,\uparrow}_{xy}$ and the two terms do not cancel for the Hall conductivity in Eq.~(\ref{eq:sigmaHupup}). Adding the two contributions $\sigma_H^{\uparrow\uparrow} + \sigma_H^{\downarrow\downarrow}$, we get the full Hall conductivity
\begin{align}
    \sigma_H(\omega) &= \frac{i e^2}{2N}\sum_\mk \nonumber \\
    &\times \left[ \frac{\epsilon_{1\mk}/E_{1\mk\uparrow} - \epsilon_{2\mk}/E_{2\mk\uparrow}}{(\omega+i0^+)^2 - (E_{1\mk\uparrow} + E_{2\mk\uparrow})^2} -\frac{\epsilon_{1\mk}/E_{1\mk\downarrow} - \epsilon_{2\mk}/E_{2\mk\downarrow}}{(\omega+i0^+)^2 - (E_{1\mk\downarrow} + E_{2\mk\downarrow})^2} \right] (\epsilon_{1\mk}-\epsilon_{2\mk})^2 \Omega^\uparrow_{1,xy}(\mk).
\end{align}
This is the same as Eq.~(20) in the main text, and completes the derivation of the Hall conductivity for equal-spin triplet, intraband only pairing. We note that, compared with Eq.~(20) in the main text, the BdG eigenenergies $E_{\mk i s}$ contain one fewer spin index. In the main text, two spin indices are instated in Eq.~(15) to express the AHR in a unified form for the two different pairing states  and Eq.~(20) is written accordingly for consistency.

\end{document}